\documentclass[11pt,a4paper]{article}
\usepackage{jheppub_kim}

\usepackage{subfigure,amsmath,amssymb ,amsfonts, latexsym}

\usepackage[notcite,color,final
                              ]{showkeys}
\definecolor{refkey}{gray}{.25}
\definecolor{labelkey}{gray}{.25}

\newcommand{\be}{\begin{equation}}
\newcommand{\ee}{\end{equation}}
\newcommand{\beqa}{\begin{subequations}\begin{eqnarray}}
\newcommand{\eeqa}{\end{eqnarray}\end{subequations}}

\title{Circularly symmetric
 solutions in three-dimensional Teleparallel, $f(T)$ and Maxwell-$f(T)$
gravity}

\author[a,b]{P.~A.~Gonz\'{a}lez}
\author[c]{Emmanuel N. Saridakis}
\author[d]{ Yerko V\'{a}squez}

\affiliation[a]{ Escuela de Ingenier\'{\i}a Civil en Obras Civiles.
Facultad de Ciencias F\'{\i}sicas y Matem\'{a}ticas, Universidad Central de
Chile, Avenida Santa
Isabel 1186, Santiago, Chile.}

\affiliation[b]{Universidad Diego Portales, Casilla 298-V, Santiago,
Chile.}

\affiliation[c]{CASPER, Physics Department, Baylor University,
Waco, TX  76798-7310, USA.}

\affiliation[d]{Departamento de Ciencias F\'{\i}sicas, Facultad de
Ingenier\'{i}a, Ciencias
y Administraci\'{o}n, Universidad de La Frontera, Avenida Francisco Salazar
01145, Casilla 54-D, Temuco, Chile.}

\emailAdd{pgonzalezm@ucentral.cl}
\emailAdd{Emmanuel$\_$Saridakis@baylor.edu}
\emailAdd{yvasquez@ufro.cl}


\abstract{We present teleparallel 3D gravity and we extract circularly
symmetric solutions, showing that they coincide with the BTZ and
Deser-de-Sitter solutions of standard 3D gravity. However, extending into
$f(T)$ 3D gravity, that is considering arbitrary functions of the torsion
scalar in the action, we obtain BTZ-like and Deser-de-Sitter-like
solutions, corresponding to an effective cosmological constant, without any
requirement of the sign of the initial cosmological constant. Finally,
extending our analysis incorporating the electromagnetic sector, we show
that  Maxwell-$f(T)$ gravity accepts deformed charged BTZ-like solutions.
Interestingly enough, the deformation in this case brings qualitatively
novel terms, contrary to the pure gravitational solutions  where the
deformation is expressed only through changes in the coefficients. We
investigate the singularities and the horizons of the new solutions, and
amongst others we show that the cosmic censorship can be violated. Such
novel behaviors reveal the new features that the $f(T)$ structure brings in
3D gravity.}

\begin{document}

\maketitle

\section{Introduction}

Although standard four-dimensional (4D) General Relativity (GR) is
believed to be the correct description of gravity at the classical level,
its quantization faces many well-known problems. Therefore,
three-dimensional (3D) gravity has gained much interest, since classically
it is much simpler and thus one can investigate more efficiently its
quantization. Amongst others, in 3D gravity one obtains the
Banados-Teitelboim-Zanelli (BTZ) black hole~\cite{BTZ}, which is a
solution to the Einstein equations with a negative cosmological constant.
This black-hole solution presents interesting properties at both classical
and quantum levels, and it shares several features of the Kerr black hole
of 4D GR~\cite{Carlip,Carlip:2005zn}. Actually it is the existence
of BTZ black holes
that makes 3D gravity a striking toy model.

Furthermore, remarkable attention was addressed recently to topologically
massive gravity, which is a generalization of 3D GR that amounts to
augment the Einstein-Hilbert action by adding a Chern-Simons gravitational
term,~\cite{deser,deser1} and thus the propagating degree of freedom is a
massive
graviton, which amongst others also admits BTZ black-hole exact
solutions. The renewed interest on  topologically
massive gravity relies on the possibility of
constructing a chiral theory of gravity at a special point of the
parameter-space, as it was suggested in~\cite{Li:2008dq}. This idea has
been extensively analyzed in the last three
years~\cite{Strominger:2008dp,Carlip:2008jk,Carlip:2008eq,Carlip:2008qh,
Giribet:2008bw,Park:2008yy,Blagojevic:2008bn,Grumiller:2008pr,
Garbarz:2008qn,Grumiller:2008qz,Grumiller:2008es,Henneaux:2009pw}, leading
to a fruitful
discussion that ultimately led to a significantly better understanding of
the model~\cite{Maloney:2009ck}. Finally, it has been shown that
3D massive gravity (where the action is given by the
Einstein-Hilbert action with a square-curvature term which gives rise to
field equations with a second order trace) admits exacts Lifshitz metrics
and black-hole solutions which are asymptotically
Lifshitz~\cite{AyonBeato:2009nh}.  

Despite the above efforts on 3D gravitational investigations, the
formulation of a quantum theory of gravity is clearly still an open
problem. Therefore, it is very interesting to study further 3D scenarios,
trying to examine their features, as an interim stage to the exploration of
4D gravity. In the present work we are interested in investigating 3D
gravity based on torsion. In particular, the so-called ``teleparallel''
equivalent of General Relativity (TEGR) \cite{ein28,Hayashi79} is an
equivalent formulation of gravity, but instead of using the curvature
defined via the Levi-Civita connection it uses the Weitzenb{\"o}ck
connection that has no curvature but only torsion. The dynamical objects in
such a framework are the four linearly independent vierbeins (these are
parallel vector fields which is what is implied by the appellations
``teleparallel''), and the advantage of this framework is that the torsion
tensor is formed solely from products of first derivatives of the tetrad.
Finally, as described in \cite{Hayashi79}, the Lagrangian density, $T$, can
then be constructed from this torsion tensor under the assumptions of
invariance under general coordinate transformations, global Lorentz
transformations, and the parity operation, along with requiring the
Lagrangian density to be second order in the torsion tensor.

In this manuscript we will present teleparallel gravity in three
dimensions in the modern language, based on the pioneering works of Kawai
\cite{Kawai1,Kawai2,Kawai3}, and we will examine its solutions and in
particular the BTZ black hole. Note that after Kawai's works, the research
on 3D gravity with torsion was performed under the light of the
unification with electromagnetism
\cite{3dgravitywithtorsion,Sousa:2003sx,garcia,Blagojevic00,Blagojevic11,
Blagojevic22,Vasquez:2009mk} or on the chiral structure
\cite{Santamaria:2011cz}, not focusing on the pure effects of
torsion which is the first goal of the present work. After this
teleparallel construction, and inspired by the fact that in four dimensions
one can generalize the theory
considering functions $f(T)$ of the torsion scalar
\cite{fT, Ferraro:2008ey,Bengochea:2008gz, Linder:2010py,Myrzakulov:2010vz,Chen001,Wu001,Bamba:2010iw,
Dent001,Zheng:2010am,Bamba:2010wb,Yerzhanov:2010vu,Yang:2010ji,
Wu:2010mn,Bengochea001,Wu:2010xk,Zhang:2011qp,Ferraro001,Cai:2011tc,
Chattopadhyay001,Sharif001,Wei001,Ferraro002,Boehmer004,Wei005,
Capozziello006,Wu007,Daouda001,Daouda:2011iy,Bamba:2011pz,Geng:2011aj,
Wei:2011yr,Geng:2011ka,Xu:2012jf,Iorio:2012cm,Wang:2011xf,Miao003,
Wei:2011aa}, we
extend our
analysis in 3D $f(T)$-gravity, too. This approach has been partially
followed in \cite{Ferraro:2011ks}, and such an investigation may be
helpful in a twofold way, that is it can be enlightening both for 3D
gravity, since new features are induced by the $f(T)$ structure, as well as
for $f(T)$ structure itself, since the 3D framework will bring light to the usual ambiguities concerning Lorentz invariance of 4D $f(T)$ gravity.
Finally, the main subject of the present work is to extend the analysis
taking into account the electromagnetic sector, in order to extract the
charged circularly symmetric solutions. As we will see, these solutions
are qualitatively different than the charged BTZ solutions of 3D General
Relativity.

 The plan of the work is as follows:  In section \ref{TEGR}, we present a
brief review of Teleparallel Equivalent to General Relativity (TEGR) in
four dimensions. In
section \ref{Tel3D}, we present the teleparallel 3D gravity and we
extract BTZ solutions, while in section \ref{3DfT} we formulate the 3D
$f(T)$ gravity, examining also circularly symmetric exact solutions. In
section \ref{3DMfT} we extend our analysis to 3D Maxwell-$f(T)$
gravity and we extract charged static black-hole solutions. Finally, in
section \ref{conclusions} we discuss the physical implications
of the results.

\section{Teleparallel Equivalent to General Relativity (TEGR)}
\label{TEGR}

In this section we briefly review TEGR in
four dimensions. Thus, our notation is
as follows: Greek indices $\mu, \nu,$... run over all coordinate space-time
0, 1, 2, 3, lower case Latin indices (from the middle of the alphabet) $i,
j, ...$ run over spatial coordinates 1, 2, 3, capital Latin indices $A, B,
$%
... run over the tangent space-time 0, 1, 2, 3, and lower case Latin
indices
(from the beginning of the alphabet) $a,b, $... will run over the tangent
space spatial coordinates 1, 2, 3.

As we stated in the Introduction, the dynamical variable of the
``teleparallel'' gravity is the vierbein field ${\mathbf{e}_A(x^\mu)}$.
This
forms an orthonormal basis for the tangent space at each point $x^\mu$ of
the manifold, that is $\mathbf{e} _A\cdot\mathbf{e}_B=\eta_{AB}$, where $%
\eta_{AB}=diag (1,-1,-1,-1)$. Furthermore, the vector $\mathbf{e}_A$ can be
analyzed with the use of its components $e_A^\mu$ in a coordinate basis,
that is $\mathbf{e}_A=e^\mu_A\partial_\mu $.

In such an construction, the metric tensor is obtained from the dual
vierbein
as 
\begin{equation}  \label{metrdef}
g_{\mu\nu}(x)=\eta_{AB}\, e^A_\mu (x)\, e^B_\nu (x)~.
\end{equation}
Contrary to General Relativity, which uses the torsionless Levi-Civita
connection, in the present formalism ones uses the curvatureless Weitzenb%
\"{o}ck connection \cite{Weitzenb23}, whose torsion tensor reads 
\begin{equation}  \label{torsion2}
{T}^\lambda_{\:\mu\nu}=\overset{\mathbf{w}}{\Gamma}^\lambda_{ \nu\mu}-%
\overset{\mathbf{w}}{\Gamma}^\lambda_{\mu\nu} =e^\lambda_A\:(\partial_\mu
e^A_\nu-\partial_\nu e^A_\mu)~.
\end{equation}
Finally, the contorsion tensor, which equals the difference between
Weitzenb%
\"{o}ck and Levi-Civita connections, is defined as 
$
K^{\mu\nu}_{\:\:\:\:\rho}=-\frac{1}{2}\left(T^{\mu\nu}_{ \:\:\:\:\rho}
-T^{\nu\mu}_{\:\:\:\:\rho}-T_{\rho}^{\:\:\:\:\mu\nu}\right)$, and it
proves useful to define 
$
S_\rho^{\:\:\:\mu\nu}=\frac{1}{2}\left(K^{\mu\nu}_{\:\:\:\:\rho}
+\delta^\mu_\rho \:T^{\alpha\nu}_{\:\:\:\:\alpha}-\delta^\nu_\rho\:
T^{\alpha\mu}_{\:\:\:\:\alpha}\right)$.

In summary, in the present formalism all the information concerning the
gravitational field is included in the torsion tensor ${T}
^\lambda_{\:\mu\nu} $. 
Using the above quantities one can define the simplest form of the
``teleparallel Lagrangian'', which is nothing else than the torsion scalar,
as \cite{Maluf:1994ji,Arcos:2005ec} {\small {\ 
\begin{equation}  \label{telelag}
\mathcal{L}=T\equiv
S_\rho^{\:\:\:\mu\nu}\:T^\rho_{\:\:\:\mu\nu}=\frac{1}{4}%
T^{\rho \mu \nu}T_{\rho \mu \nu}+\frac{1}{2}T^{\rho \mu \nu}T_{\nu \mu
\rho}-T_{\rho \mu}^{\ \ \rho}T_{\ \ \ \nu}^{\nu \mu}~.
\end{equation}%
}}  
Thus, the  simplest action of teleparallel gravity
reads: 
\begin{eqnarray}  \label{action0}
S = \frac{1}{2 \kappa}\int d^4x e \left(T+\mathcal{L}_{m}\right),
\end{eqnarray}
where $\kappa =8 \pi G$, $e = \text{det}(e_{\mu}^A) = \sqrt{-g}$ and $%
\mathcal{L}_{m}$ stands for the matter Lagrangian. We mention here that the
Ricci scalar $R$ and the torsion scalar $T$ differ only by a total
derivative \cite{Weinberg:2008}.

Variation of the action (\ref{action0}) with respect to the vierbein gives
the equations of motion 
\begin{eqnarray}\label{eom}
e^{-1}\partial_{\mu}(eS_{A}{}^{\mu\nu})
-e_{A}^{\lambda}T^{\rho}{}_{\mu\lambda}S_{\rho}{}^{\nu\mu}
-\frac{1}{4}e_{A}^{\nu
}T
= 4\pi Ge_{A}^{\rho}\overset {\mathbf{em}}T_{\rho}{}^{\nu}~,
\end{eqnarray}
where the mixed indices are used as in $S_A{}^{\mu\nu} =
e_A^{\rho}S_{\rho}{}^{\mu\nu}$. Note that the tensor
$\overset{\mathbf{em}}{T%
}_{\rho}{}^{\nu}$ on the right-hand side is the usual energy-momentum
tensor. These equations are exactly the same as those of GR for every
geometry choice, and that is why the theory is called ``Teleparallel
Equivalent to General Relativity''.

\section{3D Teleparallel  Gravity}
\label{Tel3D}

\subsection{The Model}

In this subsection we review teleparallel 3D gravity and we explore its
properties. Although the first investigations on the subject were
performed by Kawai almost twenty years ago \cite{Kawai1,Kawai2,Kawai3}, in
the following we provide the corresponding formulation using the
language of the modern literature on the subject.

As it is known, in standard 3D gravity one is inspired by the
standard 4D GR, writing:
\begin{equation}  \label{action3DR}
S = \frac{1}{2 \kappa}\int d^3x e \left(R-2\Lambda\right)~,
\end{equation}
where $\kappa$ is the three-dimensional gravitational constant, $R$ is the
Ricci scalar in 3 dimensions and $\Lambda$ the cosmological
constant. Thus, in teleparallel 3D gravity we start with the
action 
\begin{eqnarray}  \label{action}
S = \frac{1}{2 \kappa}\int d^3x e \left(T-2\Lambda\right)~,
\end{eqnarray}
where $T$ is the torsion scalar given by (\ref{telelag}), but in 3
dimensions, since the vierbeins and the metric are now three-dimensional
(the vierbeins are now a triad field instead of a tetrad one). Therefore,
in the following all the conventions that were described in the beginning
of section \ref{TEGR} run to one dimension less.

As usual it is convenient to consider the spacetime coordinates to be
$x^{\mu
}=t,r,\phi$. Thus, the torsion $T^{a}$ will simply be $T^{a}=de^{a}$. Let
us first see what the Lagrangian of teleparallel 3D gravity could be.
The more general quadratic Lagrangian in the torsion, written in
differential forms for the vielbein 1-form $e^{a}$, and under the
assumption of zero spin-connection, is given by
\cite{Muench:1998ay,Itin:1999wi} 
\begin{equation}  \label{action2}
S=\frac{1}{2 \kappa}\int \left( \rho_{0} \mathcal{L}_{0}+ \rho_{1}
\mathcal{L%
}_{1}+ \rho_{2} \mathcal{L}_{2}+\rho_{3} \mathcal{L}_{3}+ \rho_{4}
\mathcal{L%
}_{4}\right)~,
\end{equation}%
where $\rho_i$ are parameters and
\begin{equation}
\mathcal{L}_{0}= \frac{1}{4}e^{a} \wedge \star e_a~,\quad \mathcal{L}_{1}=de^{a}
\wedge \star de_{a}~,\quad \mathcal{L}_{2}= (de_{a} \wedge \star e^a)
\wedge \star (de_b \wedge e^b)~,\nonumber
\end{equation}
\begin{equation}
\mathcal{L}_{3}=(de^{a} \wedge e^{b}) \wedge \star (de_{a} \wedge
e_{b})~,\quad \mathcal{L}_{4}= (de_{a} \wedge \star e^b) \wedge
\star (de_b \wedge e^a)~,
\end{equation}
with $\star $ denoting the Hodge dual operator and $\wedge$ the wedge
product. The coupling constant $\rho_{0}=-\frac{8}{3} \Lambda$ represents
the cosmological constant term, and moreover since $\mathcal{L}_{3}$ can be
written completely in terms of $\mathcal{L}%
_{1}$, in the following we set $\rho_{3}=0$ \cite{Muench:1998ay}.  

Action (\ref{action2}) can be written in a more convenient form as
\begin{equation}
\label{actiontele0}
S=\frac{1}{2\kappa} \int \left (T -2\Lambda   \right )\star 1~,
\end{equation}
where $\star1=e^{0} \wedge e^{1} \wedge e^{2}$, and the torsion
scalar $T$ is given by
\begin{equation}  \label{scalartorsion}
T= \star \left[\rho_{1}(de^{a} \wedge \star de_{a})+\rho_{2}(de_{a} \wedge
e^a) \wedge \star (de_b \wedge e^b)+\rho_{4}(de_{a} \wedge
e^b) \wedge \star (de_b \wedge e^a) \right]~.
\end{equation}

Expanding this expression in terms of its components it is easy to obtain
the following relation 
\begin{equation}
\label{scalartorsionrho}
T=\frac{1}{2} (\rho_{1}+\rho_{2}+\rho_{4})T^{abc}T_{abc}+%
\rho_{2}T^{abc}T_{bca}-\rho_{4}T_{a}^{ac}T^{b}_{bc}~,
\end{equation}
(this is the same to the one in \cite{Kawai1,Kawai2,Kawai3} but with
different definitions of the corresponding constants).
Therefore, we straightforwardly see that for 
$\rho_{1}=0$, $\rho_{2}=-\frac{1}{2}$
and $\rho_{4}=1$ the above expression coincides with (\ref{telelag}) in
3D, namely
\begin{equation}
\label{actiontel3D}
T=\frac{1}{4}T^{abc}T_{abc}-\frac{1}{2}T^{abc}T_{bca}-T_{a}{}^{ac}T^{b}{}_{
bc}~.
\end{equation}
Finally, variation of the action (\ref{actiontele0}) with respect to the
vierbein triad provides the following field equations:
\begin{eqnarray}\label{fieldequations}
&&\delta \mathcal{L} =\delta e^{a}\wedge \left\{\left\{\rho
_{1}\left[2d\star de_{a}+i_{a}(de^{b}\wedge \star
de_{b})-2i_{a}(de^{b})\wedge
\star de_{b}\right]
\right.\right.\nonumber
\\ && \ \ \ \ \ \ \ \ \ \ \ \ \ \ \ \ \ \
+\rho _{2}\left\{-2e_{a}\wedge d\star (de^{b}\wedge
e_{b})
+2de_{a}\wedge \star (de^{b}\wedge e_{b})+i_{a}\left[de^{c}\wedge
e_{c}\wedge
\star (de^{b}\wedge e_{b})\right]\right.\nonumber\\
&&\ \ \ \ \ \ \ \ \ \ \ \ \ \ \ \ \ \ \ \ \ \, \ \ \ \ \ \left.
-2i_{a}(de^{b})\wedge e_{b}\wedge \star
(de^{c}\wedge e_{c})\right\}
 \nonumber\\
&& \ \ \ \ \ \ \ \ \ \ \ \ \ \ \ \ \ \ +\rho_{4}\left\{-2e_{b}\wedge
d\star (e_{a}\wedge de^{b})+2de_{b}\wedge \star
(e_{a}\wedge de^{b})+i_{a}\left[e_{c}\wedge de^{b}\wedge \star
(de^{c}\wedge
e_{b})\right]\right.\nonumber \\
&&\ \ \ \ \ \ \ \ \ \ \ \ \ \ \ \ \ \ \ \ \ \, \ \ \ \ \ \left.\left.
 -2i_{a}(de^{b})\wedge e_{c}\wedge \star (de^{c}\wedge
e_{b})\right\}\right\}\nonumber\\
&&\ \left. \ \ \ \ \ \ \ \ \ \ \ \ \ \ \ \
-2\Lambda \star e_a  \right\}
=0~,
\end{eqnarray}
where $i_a$ is the interior product and for generality we have kept the
general
coefficients $\rho_i$.

\subsection{Circularly symmetric Solutions}

We are interesting in circularly symmetric solutions of the above
constructed 3D teleparallel gravity. Since for the moment  we neglect
the electromagnetic sector focusing on the gravitational features of the
theory, we consider a metric ansatz of the form
\begin{equation}
  \label{metric}
ds^2=N^2dt^2-N^{-2}dr^2-r^2(d\phi+N_{\phi}dt)^2~,
\end{equation}
where $N$ and $N_{\phi}$ are  the lapse and shift functions respectively.
We mention that for the purpose of the present work we use a different
ansatz than the corresponding one in  \cite{Kawai1,Kawai2,Kawai3}.
Such an $SO(2)$ symmetric metric arises
from the following triad field up to a Lorentz transformation:
\begin{equation}
\label{BTZvierbeins}
e^{0}=Ndt~, \quad e^{1}=N^{-1}dr~, \quad e^{2}=r(d\phi+N_{\phi}dt)~.
\end{equation}
We stress here that for a linear-in-$T$ 3D of 4D teleparallel
gravity, the metric is related to the vierbeins in a simple way,
and thus relation (\ref{BTZvierbeins}) is a
safe result of (\ref{metric}). Note that this property holds for every
coefficient choice in (\ref{scalartorsionrho})
\cite{Hayashi79,Kawai1,Kawai2,Kawai3}, however only for the choice
(\ref{actiontel3D}) teleparallel gravity coincides with the
usual curvature-formulation of General Relativity.

Now, replacing the vierbein in the field equation (\ref{fieldequations}),
we obtain the following separate equations:
\begin{equation}
\left (Nr\frac{d^{2}N_{\phi}}{
dr^{2}}+3N\frac{dN_{\phi}}{dr} \right
)\left(\rho_{1}+\rho_2+\rho_4\right)=0~,
\end{equation}
\begin{equation}
-T+2 \rho_{1} \left
(N\frac{d^{2}N}{dr^{2}}+\frac{N}{r} \frac{dN}{dr}-\frac{N^{2}%
}{r^{2}} \right )-2\Lambda=0~,
\end{equation}
\begin{eqnarray}
&&2\left \{\rho_{1}\frac{dN_{\phi}}{dr} \left (-r%
\frac{dN}{dr}+N \right )-\rho_{2} \left [\frac{dN_{\phi}}{dr} \left (N+2r%
\frac{dN}{dr} \right )+rN\frac{d^{2}N_{\phi}}{dr^{2}} \right]
\right\}\nonumber \\
&&+2\rho_{4}\frac{dN_{\phi}}{dr} \left (N-r
\frac{dN}{dr} \right )-2 \Lambda -T=0~,
\end{eqnarray}
\begin{eqnarray}
&&2\left\{\rho_{1}
\left[2\frac{N}{r}
\frac{dN}{dr}-\left(\frac{dN}{dr}\right)^{2}-\left(\frac{N}{r}\right)^{2}
\right]+2\left(\rho_1+\rho_2+\rho_4\right)\left(r\frac{dN_{\phi}}{dr}
\right)^{2}\right\}\nonumber \\
&&+2\rho_{4}\left[-N\frac{d^{2}N}{dr^{2}}
-\left(\frac{dN}{dr}\right)^{2}+\frac{N}{r}\frac{dN}{dr}\right]
-2\Lambda-T=0~,
\end{eqnarray}
\begin{equation}
T+2\Lambda=0~.
\end{equation}
Therefore, one can extract the general solutions of these equations
resulting in the lapse and shift functions of the form:
\begin{eqnarray}
&&N_{\phi}( r )=-\frac{\tilde{J}}{2r^2}~,\nonumber\\
&&N(r)=Ar+\frac{B}{r}~,
\label{sol111}
\end{eqnarray}
with the integration constants $A$ and $B$ given as
\begin{equation}
A^{2}= \frac{-\Lambda}{( \rho_{4}-\rho_{1})}~, \quad
B^{2}=\frac{\tilde{J}^{2}
(\rho_{1}+\rho_{2}+\rho_{4})}{2(\rho_{1}+\rho_{4})}~,
\end{equation}
where $\tilde{J}$ is a constant. 
These solutions coincide with those obtained in \cite{Ferraro:2011ks} in
the case $\rho_1=0$, $\rho_2=-1/2$ and $\rho_4=1$. Additionally,
the horizons of the aforementioned circular solution read just
$r_{\pm}^2=-B/A$.
The above metric is similar to the extremal BTZ metric
of 3D General Relativity, which reads \cite{BTZ}:
\begin{equation}
N=\frac{r}{l}-\frac{4GMl}{r}~,\quad
N_{\phi}=-\frac{4GJ}{r^2}~,\quad
J=\pm Ml~,
\label{BTZExtremal}
\end{equation}
where the two constants of integration $M$ and $J$ are the usual conserved
charges associated with asymptotic invariance under time displacements
(mass) and rotational invariance (angular momentum)
respectively,  given by flux integrals
through a large circle at spacelike infinity, and $-1/l^2$ is the
cosmological constant \cite{BTZ}.

In order to see the similarity more transparently, let us for simplicity,
and without loss of generality, set $\rho_1=0$ (note that this is what is
expected for the standard teleparallel Lagrangian (\ref{actiontel3D})). In
this case (\ref{sol111}) can be re-written as
\begin{eqnarray}
&&N_{\phi}( r )=-\frac{\tilde{J}}{2r^2}~,\nonumber\\
&&N^{2}( r )=-\frac{\Lambda}{ \rho_{4}} r^{2}+\frac{(\rho_{2}+\rho_{4})}{2
\rho_{4}
} \frac{\tilde{J}^{2}}{r^{2}} -
\frac{\tilde{M}}{\rho_{4}}~,\label{gensoleDtelep}
\end{eqnarray}
where  $\tilde{M}$ is a  constant. Additionally,
the horizons of the aforementioned circular solution read:
\begin{equation}
r_{\pm}^2=\frac{\tilde{M}\pm\sqrt{\tilde{M}
^2+2\Lambda(\rho_2+\rho_4)\tilde{J}^2}} { -2\Lambda}~.
\end{equation}
Now we can immediately compare the above solution with the
standard BTZ solution of 3D General Relativity, which reads \cite{BTZ}:
\begin{equation}
N^2= -8GM+\frac{r^2}{l^2}+\frac{16G^2J^2}{r^2}~,\quad
N_{\phi}=-\frac{4GJ}{r^2}~.
\label{BTZ0}
\end{equation}
 If we want
solution (\ref{gensoleDtelep}) to coincide with (\ref{BTZ0}), we have
to impose the identifications that $\tilde{M}$ must be
proportional to $M$, $\tilde{J}$ proportional to $J$, and $\Lambda$
proportional to $-1/l^2$. However, apart from  $\rho_1=0$, we have to
additionally fix $\rho_4=-2\rho_2$, which up to an overall coefficient
leads exactly to
the standard teleparallel Lagrangian (\ref{actiontel3D}). This was expected
since, as we already mentioned in the previous section, it is just the form
(\ref{actiontel3D})
that leads to a complete equivalence with General Relativity. Finally, note
that in this case the torsion-scalar can be easily calculated, leading to
the constant value
\begin{equation}
\label{Tteleresult}
T=-2\Lambda~,
\end{equation}
that is the cosmological constant is the sole source of torsion.
 
At this point we have to mention that apart from the above BTZ solution,
which arises for a
negative cosmological constant $\Lambda=-1/l^2$ (under the fixing
$\rho_{1}=0$, $\rho_{2}=-\frac{1}{2}$
and $\rho_{4}=1$), we can immediately see that in the case of positive
$\Lambda$ we obtain the standard Deser-de-Sitter solution
\cite{Deser:1983tn}.

In summary, we saw that the 3D teleparallel gravity accepts the BTZ
solution (\ref{gensoleDtelep}), which coincides with that of the standard
(GR-like) 3D gravity (\ref{BTZ0}), if we use the standard teleparallel
Lagrangian (\ref{actiontel3D}). Additionally, for positive cosmological
constant we also obtain the 3D Deser-de-Sitter solution of the
standard 3D gravity. However, this coincidence with General
Relativity solutions is not the case if one goes beyond the linear
order in the torsion scalar, as we will see in the following.

\section{3D  $f(T)$  Gravity}
\label{3DfT}

\subsection{The Model}

In this section we will extend the above discussion considering arbitrary
functions of the torsion scalar $f(T)$ in the 3D gravitational action.
This procedure is inspired by the corresponding one in 4D
teleparallel gravity, where the $f(T)$ generalization exhibits many novel
features
\cite{fT, Ferraro:2008ey,Bengochea:2008gz,
Linder:2010py,Myrzakulov:2010vz,Chen001,Wu001,Bamba:2010iw,
Dent001,Zheng:2010am,Bamba:2010wb,Yerzhanov:2010vu,Yang:2010ji,
Wu:2010mn,Bengochea001,Wu:2010xk,Zhang:2011qp,Ferraro001,Cai:2011tc,
Chattopadhyay001,Sharif001,Wei001,Ferraro002,Boehmer004,Wei005,
Capozziello006,Wu007,Daouda001,Daouda:2011iy,Bamba:2011pz,Geng:2011aj,
Wei:2011yr,Geng:2011ka}, although it seems to
spoil the Lorentz invariance of
the linear theory~\cite{fTLorinv0,fTLorinv1,fTLorinv2}. Thus, we consider
an action of the form
\begin{equation}
\label{actionfT3D0}
S = \frac{1}{2 \kappa}\int d^3x e \left[T+f(T)-2\Lambda\right]~,
\end{equation}
with the torsion scalar $T$ given by (\ref{scalartorsionrho}), that is we
keep the general coefficients $\rho_i$. In differential forms the above
action can be written as:
\begin{equation}
  \label{accionfTdif}
S=\frac{1}{2 \kappa}\int \left \{ \left[f(T)+T -2\Lambda\right] \star 1 \right \}~,
\end{equation}
where now $T$ is given by (\ref{scalartorsion}). Finally, note the
difference in the various conventions in 4D $f(T)$ literature,
since some
authors replace $T$ by $f(T)$, while the majority replace $T$ by $T+f(T)$.
In this work we follow the second convention, that is the teleparallel 3D
gravity discussed in the previous section is obtained by setting $f(T)=0$.

Thus, variation with respect to the vierbein leads to
 the following field equations:
\begin{eqnarray}  \label{fieldeq}
\delta \mathcal{L} &=&\delta e^{a}\wedge
\left\{\left(1+\frac{df}{dT}\right)
\left\{\rho
_{1}\left[2d\star de_{a}+i_{a}(de^{b}\wedge \star
de_{b})-2i_{a}(de^{b})\wedge
\star de_{b}\right]\right.\right.\nonumber
\\
&&\ \ \ \ \ \ \ \ +\rho _{2}\left\{-2e_{a}\wedge d\star (de^{b}\wedge
e_{b})+2de_{a}\wedge \star
(de^{b}\wedge e_{b})+i_{a}\left[de^{c}\wedge e_{c}\wedge
\star (de^{b}\wedge e_{b})\right] \right.
\nonumber\\
&&\left. \ \ \ \ \ \ \ \  \ \ \ \ \ \ \ \,  -2i_{a}(de^{b})\wedge
e_{b}\wedge \star
(de^{c}\wedge e_{c})\right\}\nonumber\\
&&\ \ \ \ \ \ \ \ 
+\rho_{4}\left\{-2e_{b}\wedge d\star (e_{a}\wedge
de^{b})+2de_{b}\wedge \star
(e_{a}\wedge de^{b})\right.
\nonumber \\
&&\left.\left.\ \ \ \ \ \ \ \  \ \ \ \ \ \ \ \,
+i_{a}\left[e_{c}\wedge de^{b}\wedge \star (de^{c}\wedge
e_{b})\right] -2i_{a}(de^{b})\wedge e_{c}\wedge \star (de^{c}\wedge
e_{b})\right\}
\right\}\nonumber\\
&&\ \ \ \ \ \ \ \ +2\frac{d^2f}{dT^2}dT\left[\rho_1\star de_a+\rho_2
e_a\wedge
\star(de_b\wedge e^b)+\rho_4e_b\wedge \star(de^b \wedge
e_a)\right]\nonumber\\
&&\left. \ \ \ \ \ \ \ \ +\left[f(T)-T
\frac{df}{dT}\right]\wedge \star e_{a}  -2\Lambda \star e_a \right\}=0~.
\label{fieldeq000}
\end{eqnarray}

\subsection{Circularly symmetric Solutions}
\label{circsymmsol}

Similarly to the simple teleparallel case, we are interesting in
circularly symmetric solutions, and thus we consider the metric
(\ref{metric}). However, in the present case one must be careful relating
to what vierbein choice to use. In particular, as we mentioned below
relation (\ref{BTZvierbeins}), in the case of linear-in-$T$ 3D or 4D
gravity, such a simple relation between the metric and the vierbeins is
always allowed. But in the general $f(T)$ gravity in 3D or 4D this is not
the case anymore, and in general one has a more complicated relation
connecting the vierbein tetrad with the metric, with the former being
non-diagonal even for a diagonal metric \cite{fTLorinv0}. However, in the
4D cosmological investigations of $f(T)$ gravity
\cite{fT,Ferraro:2008ey,Bengochea:2008gz,
Linder:2010py,Myrzakulov:2010vz,Chen001,Wu001,Bamba:2010iw,
Dent001,Zheng:2010am,Bamba:2010wb,Yerzhanov:2010vu,Yang:2010ji,
Wu:2010mn,Bengochea001,Wu:2010xk,Zhang:2011qp,Ferraro001,Cai:2011tc,
Chattopadhyay001,Sharif001,Wei001,Ferraro002,Boehmer004,Wei005,
Capozziello006,Wu007,Daouda001,Daouda:2011iy,Bamba:2011pz,Geng:2011aj,
Wei:2011yr,Geng:2011ka,Xu:2012jf,Iorio:2012cm}, as well as in its black
hole solutions \cite{Wang:2011xf,Miao003,Wei:2011aa,Ferraro:2011ks}, the
authors still use the simple relation between the vierbeins and the
metric, as a first approach to reveal the structure and the feature of the
theory. Therefore, in the present work, in the case of 3D $f(T)$ we do
assume the simple relation between the vierbeins and the metric as a first 
investigation on this novel theory, capable of revealing the main features
of the solutions. Clearly a detailed investigation of the general vierbein
choice in 3D and 4D $f(T)$ gravity, and its relation to extra degrees of
freedom, is a necessary step for the understanding of this novel theory
\cite{inprep}.

Thus, following the above discussion we impose the vierbein ansatz
(\ref{BTZvierbeins}), and for this choice the torsion scalar
(\ref{scalartorsion}) in differential
forms reads:
\begin{equation}
  \label{Tor}
T=-\rho_{1} \left [ \left (\frac{dN}{dr} \right )^{2}+ \left (\frac{N}{r}
\right )^{2}- \left (r\frac{dN_{\phi}}{dr} \right )^{2} \right]+\rho_{2}
\left (r\frac{dN_{\phi}}{dr} \right )^{2}+\rho_{4} \left[2\frac{N}{r}
\frac{
dN}{dr}+ \left (r \frac{dN_{\phi}}{dr} \right )^{2} \right]~.
\end{equation}
Inserting this expression in the field equations (\ref{fieldeq000}), we
finally acquire the following separate equations for the metric functions:
\begin{equation} 
 \label{eq5}
\left (1+\frac{df}{dT} \right ) \left (rN\frac{d^{2}N_{\phi}}{
dr^{2}}+3N\frac{dN_{\phi}}{dr} \right
)\left(\rho_{1}+\rho_2+\rho_4\right)+Nr\frac{d^2f}{dT^2}\frac{dT}{dr}
\frac{
dN_{\phi}}{dr}\left(\rho_1+\rho_2+\rho_4\right)=0~,
\end{equation}
\begin{eqnarray}  
\label{eq1}
\nonumber &&-\left(1+\frac{df}{dT}\right)T+2 \rho_{1} \left
(1+\frac{df}{dT} \right )
\left
(N\frac{d^{2}N}{dr^{2}}+\frac{N}{r} \frac{dN}{dr}-\frac{N^{2}%
}{r^{2}} \right )\\
&&+2\frac{d^2f}{dT^2}\frac{dT}{dr}\left(\rho_1\frac{dN}{dr}-\rho_4\frac{N}{
r}
\right)N+f(T)-T\frac{df}{dT}-2\Lambda=0~,
\end{eqnarray}
\begin{eqnarray}  \label{eq2}
\nonumber &&2\left (1+\frac{df}{dT} \right ) \left
\{\rho_{1}\frac{dN_{\phi}}{dr} \left (-r%
\frac{dN}{dr}+N \right )-\rho_{2} \left [\frac{dN_{\phi}}{dr} \left (N+2r%
\frac{dN}{dr} \right )+rN\frac{d^{2}N_{\phi}}{dr^{2}} \right] \right \}\\
&&+2\rho_{4} \left (1+\frac{df}{dT} \right )\frac{dN_{\phi}}{dr} \left (N-r
\frac{dN}{dr} \right )-2\rho_2NrN_{\phi}\frac{d^2f}{dT^2}\frac{dT}{dr}=0~,
\end{eqnarray}
\begin{eqnarray}  
\label{eq4}
\nonumber
&&+2\left(1+\frac{df}{dT}\right)
 \left\{\rho_{1}
\left[2\frac{N}{r}
\frac{dN}{dr}-\left(\frac{dN}{dr}\right)^{2}-\left(\frac{N}{r}\right)^{2}
\right]+2\left(\rho_1+\rho_2+\rho_4\right)\left(r\frac{dN_{\phi}}{dr}
\right)^{2}\right\}\\
\nonumber&&+2
\rho_{4}\left(1+\frac{df}{dT}\right)
\left[-N\frac{d^{2}N}{dr^{2}}
-\left(\frac{dN}{dr}\right)^{2}+\frac{N}{r}\frac{dN}{dr}\right]\\
&&+f(T)-T\frac{df}{dT}-2\Lambda-2\frac{d^2f}{dT^2}\frac{dT}{dr}
\left(\rho_4\frac{dN}{dr}-\rho_1\frac{N}{r}\right)N-\left(1+\frac{df}{dT}
\right)T=0~,
\end{eqnarray}
\begin{equation}  \label{eq3}
\left (1+\frac{df}{dT} \right )T-\left [f(T)-T\frac{df}{dT} \right
]+2\Lambda=0~.
\end{equation}

Although the above equations seem to have a complicated form, one is able
to perform an analytical elaboration. In particular, it is worth noting
that if the form of $f(T)$ is specified, then one can use
equation (\ref{eq3}) in order to extract explicitly the value of $T$
through an algebraic equation. For instance, setting $f(T)=0$ we obtain
$T=-2\Lambda$ as expected, since it is just the simple teleparallel
result (\ref{Tteleresult}) of the previous section. For the simplest
non-trivial case which has been used in 4D $f(T)$ gravity, namely the
quadratic ansatz $f(T)=\alpha
T^{2}$, which corresponds to an ultraviolet (UV) modification of the
theory, we obtain
\begin{equation}
\label{sol1aux}
T=\frac{-1\pm \sqrt {1-24\alpha \Lambda}}{6 \alpha}~,
\end{equation}
and similarly one can find the solution for the general power-law case
 $f(T)=\alpha T^{n}$ or even for a fully general ansatz $f(T)$. Although
solving the algebraic equation (\ref{eq3}) is not possible in general, we
can straightforwardly see that the corresponding solution will not depend
on $r$, that is we can consider a form  $T=\beta$, with $\beta$ the
specific constant solution. Since $\frac{dT}{dr}=0$, equations
(\ref{eq5})-(\ref{eq3}) can be simplified significantly. Let us
investigate various solution subclasses. Observing the form of equation
(\ref{eq5}) we deduce that we have to
consider two separate cases, namely $
\rho_{1}+\rho_{2}+\rho_{4} \ne 0$ and $\rho_{1}+\rho_{2}+\rho_{4}=0$. 

\begin{itemize}

\item {\bf Case $\rho_{1}+\rho_{2}+\rho_{4} \neq 0$.}

In this case, and assuming that $f(T)\neq-T$ (which is a trivial and
unphysical case since it leads to a zero total gravitational
Lagrangian), from (\ref{eq5}) we obtain the simple equation
\begin{equation}
\label{cae111}
\frac{d^{2}N_{\phi}}{dr^{2}}=-\frac{3}{r} \frac{dN_{\phi}}{dr}~.
\end{equation}
Therefore, we acquire
 \begin{equation}\label{Nphi}
N_{\phi}(r)=-\frac{\tilde{J}}{2r^2}~,
\end{equation}
where $\tilde{J}$ is the non-trivial integration constant. Going further,
from
(\ref{eq2}) we obtain two subcases, that is  $\rho_{1}+2\rho_{2}+%
\rho_{4} \ne 0$, which proves to lead to no solution, and
$\rho_{1}+2\rho_{2}+\rho_{4} = 0$. In the later case (\ref{eq2}) is an
identity, however
  (\ref{Tor}) leads to
\begin{equation}
\label{Torinter}
N(r)=Ar+\frac{B}{r}~,
\end{equation}
with the integration constants $A$ and $B$ given as
\begin{equation}
A^{2}= \frac{\beta}{2( \rho_{4}-\rho_{1})}~, \quad
B^{2}=\frac{\tilde{J}^{2}
(\rho_{1}+\rho_{2}+\rho_{4})}{2(\rho_{1}+\rho_{4})}~,
\end{equation}
with  $\rho_{1}\neq\rho_{4}$ and $\rho_{1}\neq-\rho_{4}$,
 in order for (\ref{eq4}) to be satisfied ($T=\beta$ is the
$r$-independent solution of  (\ref{eq3})).  These solutions coincide with
those obtained in \cite{Ferraro:2011ks} in
the case $\rho_1=0$, $\rho_2=-1/2$ and $\rho_4=1$.

Comparing the obtained solution (\ref{Nphi}) and (\ref{Torinter}) with
the BTZ solution (\ref{BTZ0}), we straightforwardly observe that the
present solution is of a BTZ-like structure, however the effective
cosmological constant proportional to $A^2$ depends on the constant
$\beta$, that is on the constant solution of (\ref{eq3}) (which includes
the initial cosmological constant $\Lambda$ as well as the parameters of
the used $f(T)$ ansatz). Therefore, even if we use the standard
teleparallel Lagrangian (\ref{actiontel3D}) (that is for $\rho_{1}=0$,
$\rho_{2}=-\frac{1}{2}$ and $\rho_{4}=1$), we obtain 
\begin{eqnarray}\label{N}
&&N_{\phi}(r)=-\frac{\tilde{J}}{2r^2}~,\nonumber\\
&&N^{2}(r)=\frac{\beta}{2} r^{2}+\frac{\tilde{J}^{2}}{4r^{2}} -\tilde{M}~,
\end{eqnarray}
that is a solution that is different from the BTZ solution  (\ref{BTZ0}) of
standard 3D (GR-like) gravity, since the first term in the second
equation has a different coefficient.

We stress that the above BTZ-like solution, corresponding to an effective
cosmological constant, does not require a negative initial
cosmological constant $\Lambda$, but only   a positive $\beta$. This is
a radical difference with standard 3D gravity, and indicates the novel
features that the $f(T)$ structure induces in the gravitational theory.
Similarly, for a negative $\beta$ (and the standard torsion scalar
(\ref{actiontel3D})) we can immediately see that we obtain a 
Deser-de-Sitter-like solution \cite{Deser:1983tn} corresponding to an
effective
cosmological constant, however again we mention
that this does not require a positive initial cosmological constant.

In the specific case where
  $\rho_{1}=\rho_{4}$ we acquire
\begin{equation}
\rho_{1}=\rho_{4}~, \quad \beta=0~, \quad
B^{2}=\frac{\tilde{J}^{2}(2\rho_{1}+\rho_{2})}{4\rho_{1}}~,
\end{equation}
while for $\rho_{1}=-\rho_{4}$
 we obtain
\begin{equation}
\rho_{1}=-\rho_{4}~,\quad \tilde{J}=0~,\quad
A^{2}=-\frac{\beta}{4\rho_{1}}~.
\end{equation}

Finally, if  $\frac{dN_{\phi}}{dr}=0$ in
(\ref{cae111}), we result to $N_\phi=0$ (this integration constant is not
relevant) and to (\ref{Torinter}), but now with
\begin{equation}
A^{2}=\frac{\beta}{2(\rho_{4}-\rho_{1})}~, \quad
2B^{2}(\rho_{1}+\rho_{4})=0~,
\end{equation}
with $\rho_{2}$ unspecified.

\item {\bf Case $\rho_{1}+\rho_{2}+\rho_{4} = 0$.}

In this case equation (\ref{eq5}) is identically satisfied. Similarly to
the previous solution subclass, from equation (\ref{eq2})
we have two subcases, namely $\rho_{1}+2 \rho_{2}+\rho_{4}=0$ and
$\rho_{1}+2\rho_{2}+\rho_{4} \ne 0$.

The first subcase leads to the simpler expresions $\rho_{2}=0$
and $\rho_{1}+\rho_{4}=0$, and thus from (\ref{eq1}) we result to
\begin{equation}
N(r)=Ar+\frac{B}{r}~.
\end{equation}
Note however that now equation (\ref{Tor}) provides   only the $A$
constant:
\begin{equation}
A^{2}= \frac{ \beta}{2(\rho_{4}-\rho_{1})}~,
\end{equation}
while $B$ remains unspecified. Additionally, equations (\ref{eq2}) and
(\ref{eq4}) are trivially satisfied, and therefore $N_{\phi}$ remains
unspecified.

In the second subcase, namely $\rho_{1}+2\rho_{2}+\rho_{4} \ne 0$, we
result to the following solution 
\begin{equation}
N(r)=Ar~, \quad A^{2}=\frac{\beta}{2(\rho_{4}-\rho_{1})}~, \quad
N_{\phi}=-\frac{\tilde{J}}{2r^2}~.
\end{equation}

\end{itemize}

\section{3D Maxwell-$f(T)$ Gravity}
\label{3DMfT}

In this section we extend the previous discussion, incorporating
additionally the electromagnetic sector. In
particular, we consider an action of the
form 
\begin{equation}
  \label{accionfTMdif}
S=\frac{1}{2 \kappa}\int \left \{ \left[f(T)+T -2\Lambda \right] \star 1 \right \}+\int\mathcal{L}_{F}~,
\end{equation}
where
\begin{equation}
\mathcal{L}_{F}=-\frac{1}{2}F\wedge^{\star}F~,
\label{Mlagrangian}
\end{equation}%
corresponds to the Maxwell Lagrangian and $F=dA$, with
$A\equiv A_{\mu}dx^{\mu}$ is the electromagnetic potential 1-form. In
this case action variation leads to the following field equations:
\begin{eqnarray}  \label{fieldeq}
\delta \mathcal{L} &=&\delta e^{a}\wedge
\left\{\left(1+\frac{df}{dT}\right)
\left\{\rho
_{1}\left[2d\star de_{a}+i_{a}(de^{b}\wedge \star
de_{b})-2i_{a}(de^{b})\wedge
\star de_{b}\right]\right.\right.\nonumber
\\
&&\ \ \ \ \ \ \ \ +\rho _{2}\left\{-2e_{a}\wedge d\star (de^{b}\wedge
e_{b})+2de_{a}\wedge \star
(de^{b}\wedge e_{b})+i_{a}\left[de^{c}\wedge e_{c}\wedge
\star (de^{b}\wedge e_{b})\right] \right.
\nonumber\\
&&\left. \ \ \ \ \ \ \ \  \ \ \ \ \ \ \ \,  -2i_{a}(de^{b})\wedge
e_{b}\wedge \star
(de^{c}\wedge e_{c})\right\}\nonumber\\
&&\ \ \ \ \ \ \ \ 
+\rho_{4}\left\{-2e_{b}\wedge d\star (e_{a}\wedge
de^{b})+2de_{b}\wedge \star
(e_{a}\wedge de^{b})\right.
\nonumber \\
&&\left.\left.\ \ \ \ \ \ \ \  \ \ \ \ \ \ \ \,
+i_{a}\left[e_{c}\wedge de^{b}\wedge \star (de^{c}\wedge
e_{b})\right] -2i_{a}(de^{b})\wedge e_{c}\wedge \star (de^{c}\wedge
e_{b})\right\}
\right\}\nonumber\\
&&\ \ \ \ \ \ \ \ +2\frac{d^2f}{dT^2}dT\left[\rho_1\star de_a+\rho_2
e_a\wedge
\star(de_b\wedge e^b)+\rho_4e_b\wedge \star(de^b \wedge
e_a)\right]\nonumber\\
&&\left. \ \ \ \ \ \ \ \ +\left[f(T)-T
\frac{df}{dT}\right]\wedge \star e_{a} -2\Lambda \star e_a  - \epsilon_{abc}s^b e^c\right\}
\nonumber\\
&&\ \ \ \ \ \ \ \ \
+\delta A \left(d^{\ast
}F\right)=0~.
\label{maxwellfieldeq}
\end{eqnarray}
In the above relation we have defined
\begin{equation}
s ^ a=-\left(S ^ a_b-\frac{1}{2}\delta^{a}_{b}S\right)e^b~,
\end{equation} 
where
\begin{equation}
\label{elenermom}
S^a_b=-F^{ac}F_{bc}+\frac{1}{4}\delta^a_b\left(F^{cd}F_{cd}\right)~,
\end{equation}
is the energy momentum tensor for the electromagnetic field and $S=S^a_a$
its trace.
Although one could investigate solution subclasses with general coupling
parameters $\rho_i$, in the following for simplicity we restrict to the
usual case $\rho_1=0$, $\rho_2=-1/2$ and $\rho_4=1$ of
(\ref{actiontel3D}).

In order to extract the static, circularly symmetric solutions for
such a theory, we consider the diagonal ansatz 
 \begin{equation}
\label{BTZvierbeins2}
e^{0}=Ndt~, \quad e^{1}=K^{-1}dr~, \quad e^{2}=rd\phi~,
\end{equation}
which yields the usual metric form \cite{Carlip}
\begin{equation}
\label{metricc}
ds^2=N( r )^2dt^2-K( r )^{-2}dr^2 -r^2d\phi^2~,
\end{equation}
having in mind the discussion of the beginning of subsection
\ref{circsymmsol} on the alternative vierbein choices corresponding to the
same metric.
Concerning the  electric sector of electromagnetic 
2-form we assume   \cite{Blagojevic11} 
\begin{equation}
F=E_re^{0}\wedge e^{1}+E_{\phi}e^{2}\wedge e^{0}~, \label{F}
\end{equation}
where $E_r$ and $E_{\phi}$ are the radial and the azimuthal electric field
respectively.
Contracting the electromagnetic tensor with itself we obtain
 the electromagnetic invariant
\begin{equation}
F_{ab}F^{ab}=-2(E_r^{2}+E_{\phi}^{2})~,
\end{equation}
and thus we extract the Maxwell energy momentum tensor
\begin{equation}
S_{b}^{a}=\left(
\begin{array}{ccc}
\frac{1}{2}(E_r^{2}+E_{\phi}^{2}) & 0 & 0 \\
0 & \frac{1}{2}(E_r^{2}-E_{\phi}^{2}) & -E_rE_{\phi} \\
0 &- E_rE_{\phi} & \frac{1}{2}(-E_r^{2}+E_{\phi}^{2})%
\end{array}%
\right) ~,  \label{stress}
\end{equation}%
and its trace:
\begin{equation}
S=\frac{1}{2}(E_r^2+E_{\phi}^2)~.  \label{trace2}
\end{equation}

Inserting the above ansatzes in the field equations
(\ref{maxwellfieldeq}), we finally obtain
\begin{eqnarray}
\label{firsteq}
&&T-f(T)+2T\frac{df}{dT}+2\Lambda
+\frac{1}{2}\left(E_r^{2}-E_{\phi}^{2}\right)=0~, 
\label{maxwell1}\\
\label{maxwell2}
&&\left[1+\frac{df}{dT}\right]\left(-\frac{2K
}{r}\frac{dK}{dr}+\frac{2K^2}{Nr}\frac{dN}{dr}
\right)-2\frac{d^2f}{dT^2}\frac{K^2}{r}\frac{dT}{dr}-E_{\phi}^2=0~,\ \ \ \
\ \ \ \ \ \ \ \ \ \ \ \ \ \ \ \ \ \ \ \ \ \  \ \ \ \ \ \ \ \ 
\end{eqnarray}
\begin{equation}
\label{maxwell3}
\left[1+\frac{df}{dT}\right]\left(-2\frac{K}{N}\frac{dK}{dr}\frac{dN}{
dr } -2\frac{K^2}{N}\frac{d^2N}{dr^2} +\frac{
2K^2}
{Nr}\frac{dN}{dr}\right)-2\frac{d^2f}{dT^2}\frac{K^2}{N}\frac{dT}{dr}\frac{
dN}{dr} +E_r^2-E_{\phi} ^2=0~, 
\end{equation}
along with
\begin{eqnarray}
\label{electric}
&&E_rE_{\phi}=0~,
\\
\label{fieldE}
&&d^{\ast}F=0~, 
\end{eqnarray}
where
\begin{equation}
T=\frac{2K( r )^2N'( r )}{rN( r )}~.
\end{equation}

A first observation is that, contrary to the simple $f(T)$ case of the
previous section where the torsion scalar $T$ was a constant, in the
present case $T$ has in general an $r$-dependence, which disappears for a
zero electric charge. Such a behavior reveals the new features that
are brought in by the richer structure of the addition of the
electromagnetic sector. 

Furthermore, form  (\ref{electric}) we deduce that either $E_{\phi}=0$ or
$E_r=0$, that is we cannot have simultaneously non-zero radial and
azimuthal electric field. This is an interesting result, since it shows
that the known no-go theorem of 3D GR-like gravity
\cite{Cataldo:2002fh,Blagojevic00}, that configurations with two
non-vanishing components of the Maxwell field are dynamically not allowed,
holds in 3D $f(T)$ gravity too. At this point one could ask whether this
result is accidental, holding only for the diagonal vierbein choice
(\ref{BTZvierbeins2}). However, as we show in appendix \ref{nogogeneral},
a general vierbein choice, although it will change the solution structure,
it will still lead to the above no-go theorem, which is thus a general
result. So let us investigate these two electric-field cases separately.

\subsection{Absence of azimuthal electric field}

In the case $E_{\phi}=0$, that is in the absence of azimuthal electric
field, equation (\ref{fieldE}) leads to  
\begin{equation}
E_r=\frac{Q}{r}~,
\end{equation}
where $Q$ is an integration constant, that as usual coincides with
the electric charge of the circular object (black hole). In order to
proceed, we will consider Ultraviolet (UV) and Infrared (IR) corrections to
$f(T)$ gravity respectively.

\subsubsection{UV modified 3D gravity}
\label{UVmod}

In order to examine the modifications on the circular solutions caused by
UV modifications of  3D gravity we consider a representative ansatz of the
form  $f(T)=\alpha T^{2}$. Thus, for $\alpha\neq0$ equation
(\ref{maxwell1}) gives: 
\begin{equation}
\label{aux11}
T=\frac{-1\pm\sqrt{1-12\alpha\left(2\Lambda+\frac{Q^2}{2r^2}\right)}}{
6\alpha
}~,
\end{equation}
with the upper and lower signs corresponding to the positive and negative
branch solutions respectively.\footnote{\label{footonotebranch}We mention
here that if
$\alpha=0$ then (\ref{maxwell1}) becomes linear and it has only
one solution, which is given by the $\alpha\rightarrow0$ limit of the
positive branch of (\ref{aux11}), namely $T(r)=-Q^2/(2r)$. In this case
teleparallel gravity is restored and the corresponding solutions coincide
with the BTZ ones of General Relativity.} Choosing for simplicity and
without loss of generality
that $\Lambda=0$, we obtain the solution
\begin{equation}
\label{Tcase1}
T( r )=\frac{-1\pm\sqrt{1-6\frac{\alpha Q^2}{r^2}}}{6\alpha}~,
\end{equation}
corresponding to
\begin{eqnarray}\label{array1}
 &&N(r)^2 =\frac{1}{108}\left\{-\frac{1}{\alpha}\left\{
r^2\mp P(r)[12\alpha
Q^2+r^2]+36\alpha 
Q^2 \ln r \pm18\alpha Q^2 \ln
\{r[1+P(r)]\}\right\}+const\right\}~,\nonumber
\\
&&K(r)^2= N(r)^2\left[\frac{2}{3}\pm\frac{1}{3}P( r )
\right]^{-2}~,\label{case1}
\end{eqnarray}
where 
\begin{equation}
P( r )=\sqrt{1-\frac{6\alpha Q^2}{r^2}}~.
\end{equation}
As we observe, in the zero-electric-charge limit, the above solutions
coincide with those of (\ref{sol1aux}),(\ref{N}), with $\tilde{J}=0$,
$\Lambda =0$ and  $f\left( T\right) =\alpha T^{2}$, as expected.

Let us now investigate the singularities and the horizons of the above
solutions. The first step is to find at which $r$ do the functions $N(r)$
and $K(r)$ become zero or infinity. However, since these singularities may
correspond to just coordinate singularities, the usual procedure is to
investigate various invariants, since if these invariants diverge at one
point they will do that independently of the specific coordinate basis. In
standard black-hole literature of the curvature-formulated gravity (either
General Relativity or its modifications) one usually studies the Ricci
scalar, the Kretschmann scalar, or other invariants constructed by the
Riemann tensor and its contractions. In teleparallel description of
gravity one needs to examine curvature invariants too, but using the
standard Levi-Civita connection, since the use of the Weitzenb{\"o}ck
connection leads to zero curvature invariants by construction. The use of
curvature invariants instead of torsion ones is also indicated by the fact
that in a realistic theory matter is coupled to the gravitational sector
through the metric and not the vierbeins (with the interesting exception
of fermionic matter), and particles follow geodesics defined by the
Levi-Civita connection. Whether one can formulate everything in terms of
the Weitzenb{\"o}ck connection and torsion invariants, as he can do with
the Levi-Civita connection and curvature invariants, is an open subject and
needs further investigation, in particular relating to the quantization
procedure (where the fundamental field, the metric or the vierbeins, should
be determined).

\begin{figure}
\begin{center}
\includegraphics[width=4.0in,angle=0,clip=true]{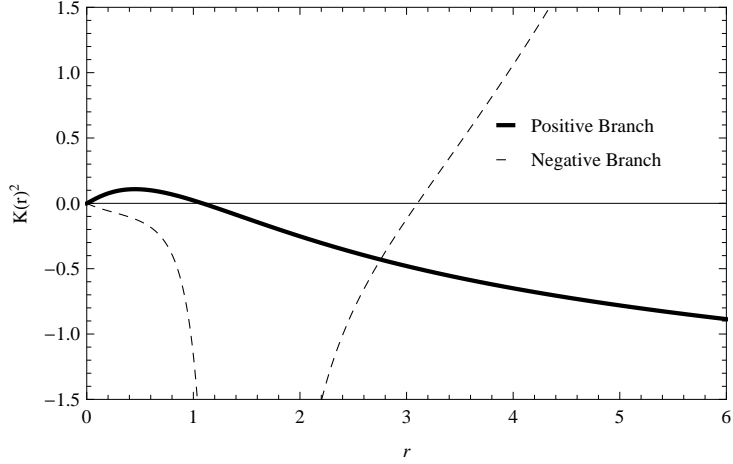}
\caption{The solution $K(r)^2$ of (\ref{case1}) as a function of
$r$, for the positive (thick curve) and negative (thin curve) branch of the
UV modified 3D Maxwell-$f(T)$ gravity, for $\alpha=-1$, $Q=1$ and
$const=-1$.}
\label{fig1}
\end{center}
\end{figure}

Thus, in order to proceed to the singularities and horizons investigation
along the above lines, we have to first solve the equation $K(r)^2=0$.
However, from the form of $K(r)^2$ in (\ref{case1}) we can clearly see
that in general this is a transcendental equation, whose roots cannot be
obtained analytically apart from the root at $r=0$. Thus, in the following
we will examine numerically a specific case, choosing without loss of
generality $\alpha=-1$ (note that for $\alpha<0$, the torsion scalar
(\ref{Tcase1}) is always real)  $Q=1$ and $const=-1$, and in
Fig.~\ref{fig1} we depict $K(r)^2$ as a function of $r$ for both the
positive and negative branch.

\begin{figure}
\begin{center}
\includegraphics[width=4.0in,angle=0,clip=true]{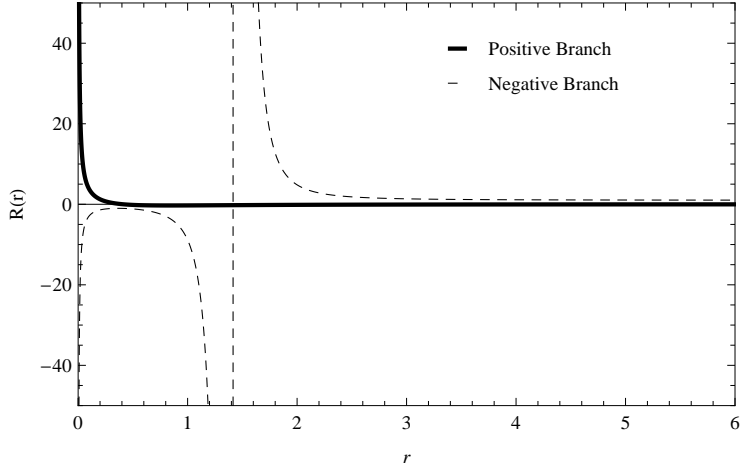}
\caption{The Ricci scalar $R( r )$ as a function of $r$, for the positive
(thick curve) and negative (thin curve) branch of the UV modified 3D
Maxwell-$f(T)$ gravity, for $\alpha=-1$, $Q=1$ and $const=-1$.}
\label{fig2}
\end{center}
\end{figure}

\begin{figure}
\begin{center}
\includegraphics[width=4.0in,angle=0,clip=true]{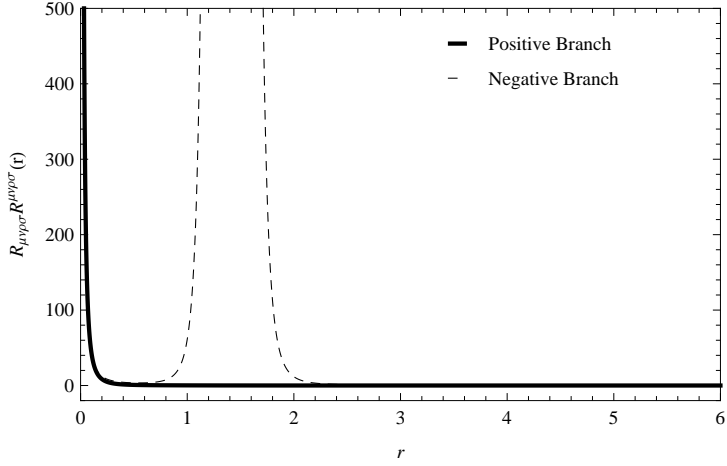}
\caption{The Kretschmann scalar $R_{\mu\nu\rho\sigma}R^{\mu\nu\rho\sigma}
(r )$ as a function of $r$, for the positive (thick curve) and negative
(thin curve) branch of the UV modified 3D Maxwell-$f(T)$ gravity, for
$\alpha=-1$, $Q=1$ and $const=-1$.}
\label{fig3}
\end{center}
\end{figure}

For the positive branch we observe that apart from the root at $r=0$,
there is another value $r=r_H$ where $K(r_H)^2=N(r_H)^2=0$.
Additionally, in Fig.~\ref{fig2} and Fig.~\ref{fig3} we respectively
depict the Ricci scalar
$R(r)$ and the Kretschmann scalar $R_{\mu\nu\rho\sigma}R^{\mu\nu\rho\sigma}
(r)$, calculated using the Levi-Civita connections (which are easily
calculated since the metric is known) as described above. As we can see,
both these invariants exhibit a physical singularity at $r=0$, however they
are regular at $r=r_H$. 

In order to ensure whether $r=r_H$ is a physical singularity or a horizon
one, we consider the Painlev\'{e}-Gullstrand coordinates \cite{Painleve, Gullstrand, Martel:2000rn, Liu:2005hj} through the
transformation
\begin{equation}
dt=d\tau +f\left( r\right) dr,
\end{equation}
with $f\left( r\right)$ a function of the radial coordinate $r$. Thus, the
metric  (\ref{metricc}) writes as
\begin{equation}\label{metricPG}
ds^{2}=N\left( r\right) ^{2}d\tau ^{2}+2f\left( r\right) N\left( r\right)
^{2}drd\tau -\left[\frac{1}{K\left( r\right) ^{2}}-N\left( r\right)
^{2}f\left( r\right) ^{2}\right] dr^{2}-r^{2}d\phi ^{2},
\end{equation}
and choosing $f\left( r\right) ^{2}=\frac{1}{N\left( r\right) ^{2}}\left[
\frac{1}{K\left(
r\right) ^{2}}-1\right]$  and setting $h(r)^2= N\left( r\right)
^{2}/K\left( r\right) ^{2}=[2+ P( r)]/3$  
we can write it in a flat Euclidean form
\begin{equation}\label{metricregular}
ds^{2}=N\left( r\right) ^{2}d\tau ^{2}+2h\left( r\right) \sqrt{1-K\left(
r\right) ^{2}}drd\tau -dr^{2}-r^{2}d\phi ^{2}~,
\end{equation}
which is regular at $r=r_H$. Therefore,  $r=r_H$ is a coordinate
singularity.

Furthermore, in order to examine whether $r=r_H$  is a Killing
horizon we observe that the timelike Killing vector of the metric is
$\epsilon^{\mu}\partial_{\mu}=\partial_t$ \footnote{Since none
of the metric coefficients depends on time, the manifold has a
timelike Killing vector $\partial_t$, and similarly since none of
the metric coefficients depends on $\phi$ then there exist
a spacelike Killing vector field $\partial_\phi$ (the metric is
circularly symmetric).}, with norm
$\epsilon_{\mu}\epsilon^{\mu}=g_{tt}=N( r )^2$ which vanishes at $r=r_H$.
Outside the horizon the Killing vector field is spacelike, while inside it
is timelike and thus it corresponds to a null hypersurface, that is a
cosmological Killing horizon.
  
For the negative branch  $K(r)^2=0$ at $r_{s}=0$ and at  $r=r_H$, however
between these two values, and contrary to the positive branch, we have a
third singularity where $|K(r_{inf})^2|\rightarrow\infty$, and according
to
 (\ref{case1}) this happens when $P\left(r_{inf}\right) =2$, that is at
$r_{inf}=\sqrt{-2\alpha Q^2}$ (in our specific numerical example
$r_{inf}=\sqrt{2}$). Now, from
Fig.~\ref{fig2} and Fig.~\ref{fig3} we
observe that the curvature invariants diverge at $r_s=0$ and at $r_{inf}$,
while they remain regular at $r=r_H$. Therefore, we conclude that $r_{inf}$
is a physical singularity while $r=r_H$ is a Killing horizon, corresponding
to an event horizon since the Killing vector field is
timelike outside the horizon and it is spacelike inside. Similarly, to the
previous case, in the Painleve-Gullstrand coordinates the metric is regular
at the event horizon, and therefore we have a coordinate singularity.
Finally, in the specific numerical example of
Figures  \ref{fig1}, \ref{fig2} and \ref{fig3} we observe that
$r_{inf}<r_H$, and thus the singularity will be shielded by the
horizon. However, in general we can have $r_{inf}>r_H$, that is a naked
singularity, and this happens when   
\begin{equation}\label{restrictiona}
const-18Q^{2}-18Q^{2}\ln \left( \sqrt{-2\alpha Q^2}\right) +18Q^{2}\ln
3 >0~, 
\end{equation}
as can be seen by investigating the singularities and the root structure
of $N^2( r )$ in (\ref{case1}) (examining its first and second
derivatives), and calculating also $N^2(r_{inf})$.

Going beyond the above specific numerical example, we may still obtain 
analytical expressions for the horizon $r_H$, in specific limits. In
particular, if $\left |6\alpha Q^{2}\right |\ll r_H^{2}$  for the positive
branch we obtain
\begin{equation}
r_{H}\approx 2^{-\frac{1}{3}}e^{\frac{1}{6}\left( 1+\frac{const}{
9Q^{2}}\right) }~.
\end{equation}
Similarly, for the negative branch we acquire 
\begin{equation}
r_{H}\approx 2e^{\frac{1}{2}\left( -1+\frac{const}{9Q^{2}}\right) }e^{-\frac{
1}{2}W_{0}\left( \frac{8}{9\alpha Q^{2}}e^{-1+\frac{const}{9Q^{2}}}\right) }~,
\label{ec1}
\end{equation}
where $W_{0}$ stands for the main branch of the Lambert function
that is single-valued since $W_{0}\left( x\right) \geq -1$ \cite{Hilbert}
(note that this is real if the argument of the Lambert
function is greater than $-1/e$, that is if 
$\frac{8}{9\alpha Q^{2}}>-e^{-\frac{const}{9Q^{2}}}$).

We close this paragraph by mentioning that in the scenario at hand the
physical singularities are not always shielded by the horizon. Thus,
the cosmic censorship does not always hold for 3D Maxwell-$f(T)$
gravity in the absence of azimuthal electric field.

\subsubsection{IR modified 3D gravity}
\label{IRmod}

In order to examine the modifications on the circular solutions caused by
IR modifications of  3D gravity we consider a representative ansatz of the
form  $f(T)=\alpha T^{-1}$. In this case equation (\ref{maxwell1}) gives:
\begin{equation}
\label{Tcase2}
T( r )= -\left(\Lambda + \frac{Q^2}{4r^2}\right) \pm
\frac{1}{2}\sqrt{12\alpha+\left(2\Lambda+\frac{Q^2}{2r^2}\right)^2}~,
\end{equation}
with the upper and lower signs corresponding to the positive and negative
branch solution
respectively. Choosing for simplicity  $\Lambda=0$, we result to:
\begin{eqnarray}
 &&N(r)^2 =-\frac{Q^6}{1728\alpha
r^4}\pm\left[\frac{1}{18}-\frac{Q^4}{1728 \alpha
r^4}\right]Y(r)-\frac{1}{3}Q^2
\ln r  \pm\frac{1}{12} Q^2 \ln \left[\frac{r^2}{2Q^2+2Y(r)}\right]+const~,
\nonumber
\\
&&K( r )^2= N( r )^2\left\{1-\frac{16\alpha r^4}{[- Q^2\pm Y( r )]^2}
\right\}^{-2}~,
\label{case2}
\end{eqnarray}
where 
\begin{equation}
Y( r )=\sqrt{Q^4+48\alpha r^4}~.
\end{equation}
As we observe, in the zero-electric-charge limit, the above solutions
coincide with those of (\ref{N}), with $\tilde{J}=0$, $\Lambda =0$ and
$f(T)=\alpha T^{-1}$, as expected (in this case $T=\beta$ will be the
specific constant solution of the algebraic equation  (\ref{eq3})).

\begin{figure}[ht]
\begin{center}
\includegraphics[width=4.0in,angle=0,clip=true]{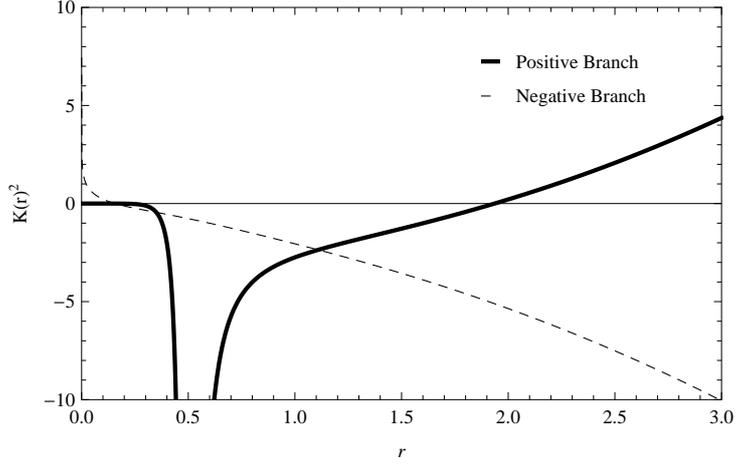}
\caption{The solution $K(r)^2$ of (\ref{case2}) as a function of $r$, for
the positive (thick curve) and negative (thin curve) branch of the IR
modified 3D Maxwell-$f(T)$ gravity,
 for   $\alpha=1$, $Q=1$ and $const=-1$. For the negative branch,
$|K(r)^2|\rightarrow\infty$ at $r=0$, which cannot be clearly seen in the
figure scale.}
\label{fig4}
\end{center}
\end{figure}

\begin{figure}
\begin{center}
\includegraphics[width=4.0in,angle=0,clip=true]{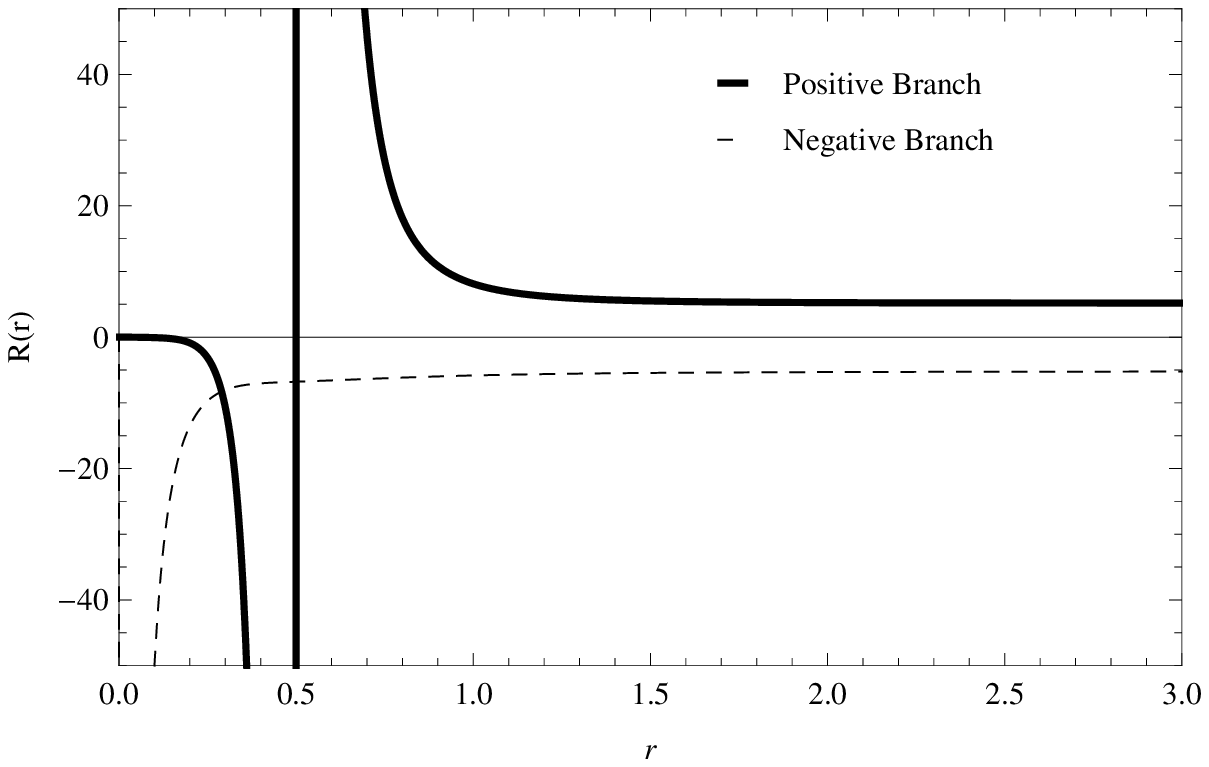}
\caption{The Ricci scalar $R (r )$ as a function of $r$, for the positive
(thick curve) and negative (thin curve) branch of the IR modified 3D
Maxwell-$f(T)$ gravity, for $\alpha=1$, $Q=1$ and $const=-1$. }
\label{fig5}
\end{center}
\end{figure}

\begin{figure}
\begin{center}
\includegraphics[width=4.0in,angle=0,clip=true]{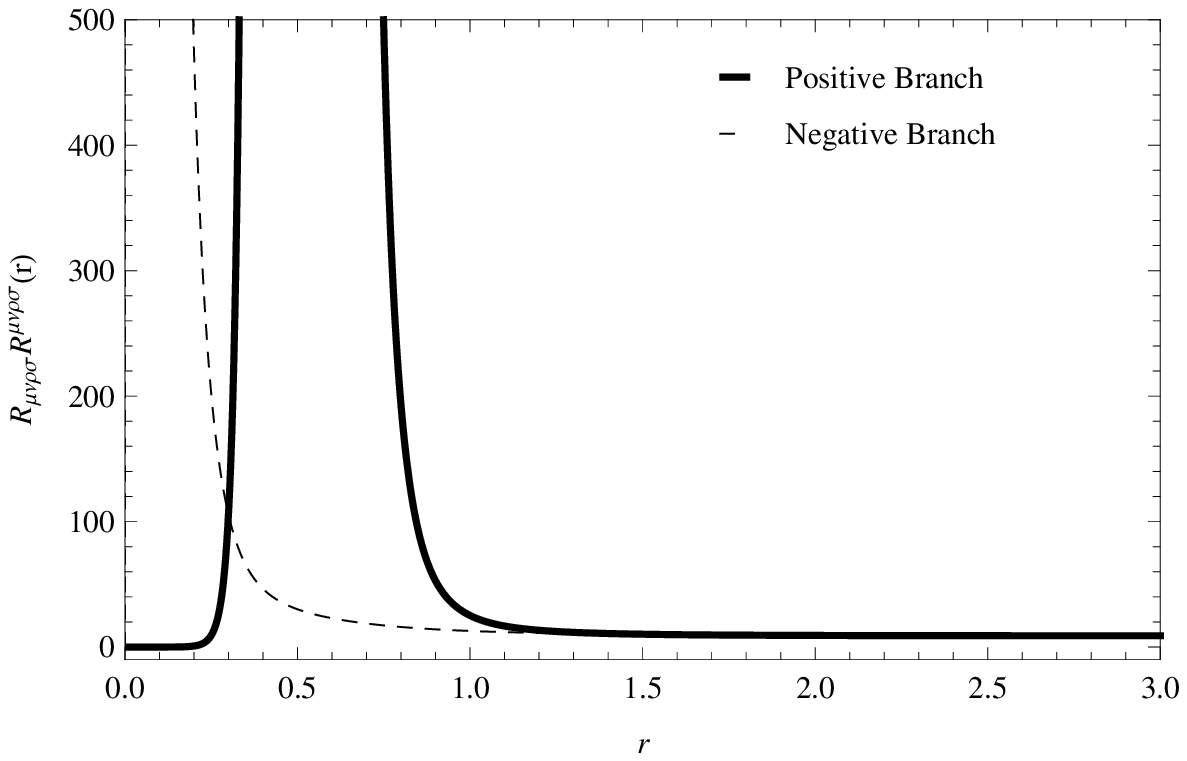}
\caption{The Kretschmann scalar $R_{\mu\nu\rho\sigma}R^{\mu\nu\rho\sigma}
(r )$ as a function of $r$, for the positive (thick curve) and negative
(thin curve) branch of the IR modified 3D Maxwell-$f(T)$ gravity, for
$\alpha=1$, $Q=1$ and $const=-1$.}
\label{fig6}
\end{center}
\end{figure}

Similarly to the previous paragraph \ref{UVmod}, we will investigate the
singularities and the horizons. Since the transcendental equation
$K(r)^2=0$, apart from the root at $r=0$, cannot be solved analytically,
in the following we examine numerically a specific result, choosing without
loss of generality  $\alpha=1$ (for $\alpha>0$, the torsion scalar is
always real), $Q=1$ and $const=-1$. Thus, in Fig.~\ref{fig4} we depict 
$K(r)^2$ as a function of $r$ for both the positive and negative
branch, while in Figures \ref{fig5} and \ref{fig6} we respectively present
the
Ricci and Kretschmann scalar.

For the positive branch we have a singularity at $r_s=0$ and one at $r_H$,
both of them corresponding to $K(r)^2=0$. However, for both of them the
curvatures scalars remain finite, while in the Painleve-Gullstrand
coordinates the metric is regular and therefore these are just coordinate
singularities. The timelike Killing vector of the metric is
$\epsilon^{\mu}\partial_{\mu}=\partial_t$ and thus its norm
$\epsilon_{\mu}\epsilon^{\mu}=g_{tt}=N( r )^2$ vanishes at $r=r_H$, since
$N(r_H)^2=0$. Outside  $r=r_H$, the Killing vector field is
timelike, while inside it is spacelike, and thus it corresponds to a null
hypersurface, that is a Killing horizon.
Additionally, we have a third singularity when
$|K(r_{inf})^2|\rightarrow\infty$, and according to
(\ref{case2}) this happens when $16\alpha
r_{inf}^4=\left[-Q^{2}+Y\left( r_{inf}\right) \right]^2$, that is at
 $r_{inf}=Q/(2\alpha ^{1/4})$ (in our
specific numerical example $r_{inf}=1/2$). As we observe from Figures
\ref{fig5} and \ref{fig6}, the Ricci and Kretschmann scalar diverge at
$r_{inf}$ too, and thus it is a physical singularity. 
Although in the specific
numerical example of Figures  \ref{fig4}, \ref{fig5} and \ref{fig6} we
observe that $r_{inf}<r_H$, and therefore the singularity will be shielded
by the horizon, in general  we can have a naked singularity if
\begin{equation}
\label{restrictionb}
108const+9Q^{2}-18Q^{2}\ln \left( \frac{1}{2\alpha ^{1/4}}Q\right)
-9Q^{2}\ln \left( 6Q^{2}\right) >0~,
\end{equation}
as can be seen from the root structure of $N^2(r)$ in (\ref{case2}).

For the negative branch we have two singularities. The first is at 
$r_{inf}=0$, corresponding to $|K(r_{inf})^2|\rightarrow\infty$ (it cannot
be clearly seen in the scale of Fig.~\ref{fig4}, but it can be
immediately verified by considering the $r\rightarrow0$ limit of
(\ref{case2})). This singularity is accompanied by a divergence in the
curvature scalars in Figures \ref{fig5} and \ref{fig6}, and thus it is a
physical and not a coordinate one. The second singularity is at $r_H$,
corresponding to $K(r)^2=0$, where the curvature scalars remain finite. 
Since in the Painleve-Gullstrand coordinates one can show that the metric
is regular at $r_H$, we deduce that it is just a coordinate singularity.
In the same lines as above, $r_H$  is a cosmological Killing horizon  which
moreover in this
specific example shields the physical singularity at $r_{inf}=0$. Note
that in the present case, independently of the parameter values we cannot
have a naked singularity.

As in previous subsection, we can go beyond the specific numerical
example and obtain analytical expressions for the horizon $r_H$, in
specific limits. In particular, if $\left |\frac{Q^4}{48\alpha}\right |\ll
r_H^4$ we obtain
\begin{equation}
r_{H}\approx \left( 2\alpha \sqrt{48}\right) ^{\mp \frac{1}{4}}e^{\frac{%
3const}{Q^{2}}}e^{-\frac{1}{2}W_{0}\left( \mp \frac{\sqrt{48\alpha }}{3Q^{2}}%
\left( 2\alpha \sqrt{48}\right) ^{\mp \frac{1}{2}}e^{\frac{6const}{Q^{2}}%
}\right) }~,
\end{equation}
where as usual the upper sign corresponds to the positive branch and the
lower sign corresponds to the negative branch. This is real
  if 
\begin{equation}
\mp \frac{\sqrt{48\alpha }}{3Q^{2}}\left( 2\alpha \sqrt{48}\right) ^{\mp 
\frac{1}{2}}e^{\frac{6const}{Q^{2}}}>-\frac{1}{e}~.
\end{equation}

Finally, similarly to the UV modification, in the present IR modified case
we observe that the physical singularities are not always shielded by the
horizon. Thus, we also verify that the cosmic censorship does not always hold for
3D Maxwell-$f(T)$ gravity in the absence of azimuthal electric field.

\subsubsection{Comparison with the BTZ-like solution in the absence of
azimuthal electric field}

Let us compare the above solutions (\ref{case1}) and
(\ref{case2}) with the charged BTZ-like solution of 3D General
Relativity in the absence of
azimuthal electric field \cite{Carlip}:
\begin{equation}
\label{chargedBTZ}
N(r)^2=K( r )^2=
-8GM+\frac{r^2}{l^2}-\frac{1}{2}Q^2\ln\left(\frac{r}{r_0}\right).
\end{equation}
As we observe,  solutions (\ref{case1}) and (\ref{case2}) correspond to a
``deformed'', charged BTZ-like solution, and they completely coincide with
it in the limit $f(T)\rightarrow0$ (that is when  $\alpha\rightarrow0$).  
Finally, as we have mentioned, in the zero
electric charge limit we re-obtain the results of the previous section.

Here we would like to stress that the deformation of the solutions
(\ref{case1}) and (\ref{case2}), comparing to the standard charged BTZ
solution (\ref{chargedBTZ}), is not of a trivial type, since we obtain
qualitatively different novel terms, corresponding to different behavior,
as it was analyzed in detail in the previous paragraphs. This was not the
case in the
pure gravitational solutions of the previous section, where the deformation
was expressed only through changes in the coefficients. Such a novel
behavior of the Maxwell-$f(T)$ theory reveals the new features that the
$f(T)$ structure brings in 3D gravity.

Despite the radical difference with the charged BTZ-like solution, away
from the circular object we obtain similarities. In particular, for the UV
modified case of paragraph \ref{UVmod}, the asymptotic behavior of the
solutions (\ref{case1}), that is at $r \to \infty$, is
\begin{equation} 
N\left( r\right) ^{2} \to -\frac{Q^2}{2}\ln r~,\ \ \ 
K\left( r\right) ^{2}\to N\left( r\right) ^{2}~,
\end{equation}
for the positive branch,
while for the negative branch we obtain
\begin{equation} 
N\left( r\right) ^{2}\to  -\frac{r^2}{54\alpha}-\frac{Q^2}{6}\ln r~,\ \ \  
K\left( r\right) ^{2}\to 9N\left( r\right) ^{2}.\
\end{equation}
As we observe the asymptotic behavior of the
negative branch coincides with that of the usual 
BTZ black hole (\ref{chargedBTZ}), while the asymptotic behavior of the
positive branch presents novel behavior.

Similarly, for the case of IR modification of paragraph \ref{IRmod}, the
asymptotic behavior of the
solutions is
\begin{eqnarray}
N\left( r\right) ^{2}\to\pm
\frac{2}{9}\sqrt{3\alpha}r^{2}-\frac{1}{3}Q^{2}\ln r~,\ \ \  
K\left( r\right) ^{2}\to \frac{9}{4}N\left( r\right) ^{2}~,\
\end{eqnarray}
with the upper and lower sign corresponding to the positive and
negative branch respectively. In this case we observe that 
the asymptotic behavior of the positive branch coincides with that of the
usual BTZ black hole (\ref{chargedBTZ}), while the asymptotic behavior of
the negative branch exhibits novel behavior.

\subsection{Absence of radial electric field}

In the case $E_r=0$, that is in the absence of radial electric
field, equation (\ref{fieldE}) leads to  
\begin{equation}
E_{\phi}=\frac{Q}{r^2}~,
\end{equation}
where $Q$ is an integration constant, that as usual coincides with
the electric charge.
This case is simpler than case  $E_{\phi}=0$ of the previous subsection,
and in particular it allows for the extraction of $N(r)$ and $K(r)$ for a
general $f(T)$, namely:
\begin{eqnarray}\label{array2}
&&N(r)= \gamma r~,
\nonumber\\
&&K(r)= \sqrt{\frac{T(r)r^2}{2}}~,
\label{noradial}
\end{eqnarray}
where $\gamma$ is an integration constant. 
Interestingly enough, since in the present case the metric in terms of the
torsion scalar has the very simple form  (\ref{noradial}), one can
calculate and express the Levi-Civita connections and then the Riemann
tensor and the curvature scalars, in terms of the torsion scalar too.
In particular, for both branches we obtain:
\begin{equation}
R\left( r\right) =3T\left( r\right) +rT^{\prime }\left( r\right)~, 
\end{equation}
\begin{equation}
R_{\mu \nu \rho \sigma }R^{\mu \nu \rho \sigma }\left( r\right) =3T\left(
r\right) ^{2}+2rT\left( r\right) T^{\prime }\left( r\right) +\frac{1}{2}
r^{2}\left[ T^{\prime }\left( r\right) \right] ^{2}~,
\label{curvT}
\end{equation}
where we mention that the scalar curvatures are defined through the
Levi-Civita connections, while the torsion scalar is defined using the
Weitzenb{\"o}ck one.

For completeness, we explicitly present the $T(r)$ solution in
the case of UV and IR modifications of 3D gravity of the previous
subsection. In the case of a UV modification of the form
$f(T)=\alpha T^{2}$
we obtain
\begin{equation}
T( r )= \frac{-1 \pm \sqrt{1-24\alpha\Lambda+\frac{6\alpha
Q^2}{r^4}}}{6\alpha}~,
\label{noradial1}
\end{equation}
with the upper and lower sign corresponding to the positive and
negative branch respectively, and we can see that for both branches  
the torsion scalar is always real if $\alpha<(24\Lambda)^{-1}$.

For the positive branch we have a singularity at $r_s=0$ where 
$N^2(r_s)=0$, and since $T(0)$ diverges, from (\ref{curvT}) we deduce that 
the Ricci and Kretschmann scalar diverge too, thus $r_s=0$ is a physical 
singularity. However, for $\Lambda>0$ we obtain a second singularity where
$K^2(r_H)=0$, 
namely at $r_H=\left(\frac{Q^2}{4\Lambda}\right)^{1/4}>0$, which due to 
(\ref{noradial}) corresponds to $T(r_H)=0$. We mention here that the fact that $K(r)$ is given by a function of the covariant quantity $T(r)$ in 
the given coordinate system, it does not mean that is will do so in another coordinate choice, since $K(r)$ is not itself a covariant quantity. Thus, 
$K^2(r_H)$ can be non-zero although  $T(r_H)=0$ still holds. In particular
\footnote{We thank an anonymous referee for his clarification on  this
subject.}, in the vicinity of $r_H$ we observe that $K^2(r)$ behaves like
$-4r_H\Lambda(r-r_H)$, neglecting higher terms in
$(r-r_H)$, and thus defining $H=r-r_H$, the metric, for $H>0$, can be
approximately written as
\begin{equation} 
ds^2=\gamma^2\left(r_H\Lambda
L^2+r_H\right)^2dt^2+dL^2-\left(r_H\Lambda
L^2+r_H\right) ^ 2 d
\phi^2~,
\end{equation}  
with $L^2=H/(r_H\Lambda)$, while for $H<0$ it becomes
\begin{equation} 
ds^2=\gamma^2\left(-r_H\Lambda
L^2+r_H\right)^2dt^2-dL^2-\left(-r_H\Lambda
L^2+r_H\right)^2d
\phi^2~, 
\end{equation}
with  $L^2=-H/(r_H\Lambda)$. As we can clearly see all
metric components are smooth functions of the coordinates around $L=0$,
and therefore $r_H$ is not a curvature singularity. However, note that the
metric changes signature at $r_H$ so it is definitely singular at $r_H$.
Additionally, by examining the Killing vector, the norm of the timelike
Killing vector $\epsilon ^{\mu }\partial _{\mu }=\partial _{t}$ is
$\epsilon _{\mu }\epsilon 
^{\mu }=N\left( r\right) ^{2}=\gamma ^{2}r^{2}$ (note that $\gamma$ can be absorbed in $dt$), and thus at  $r=r_H$ we 
 have $\epsilon _{\mu }\epsilon ^{\mu }=\gamma ^{2}r_{H}^{2}$. Therefore, 
 the norm is different from zero and we do not have 
 a Killing horizon. 
 
For the negative branch we have a physical singularity at $r_s=0$, in
which the Ricci and Kretschmann scalar diverge, however in this case there
is not any other singularity at $r>0$ (unless $\alpha=0$, but in this case
the negative branch is meaningless similarly to the case of
footnote \ref{footonotebranch}). Therefore, in this case $r_s=0$ is a naked
singularity, and  the cosmic censorship does not hold. 

In the case of an IR modification  of the form 
$f(T)=\alpha T^{-1}$
we acquire 
\begin{equation}
T( r )=
-\Lambda+\frac{Q^2}{4r^4}\pm\sqrt{3\alpha+\left(-\Lambda+\frac{Q^2}{4r^4}
\right)^2}~,
\label{noradial2}
\end{equation}
and the torsion scalar is always real if
$\alpha>-\Lambda^{2}/3$. Both branches have a physical singularity at
$r_s=0$, at which the torsion scalar  and thus the curvature
scalars in (\ref{curvT}) diverge, however both branches do not
have a horizon at $r>0$  (unless $\alpha=0$). Therefore, in this case the
singularity at $r_s=0$ is a naked one and the cosmic censorship does not
hold. 

Let us compare the above solutions  (\ref{noradial}),(\ref{noradial1}) 
and (\ref{noradial}),(\ref{noradial2})  with the charged BTZ-like solution
in the absence of radial electric field \cite{Carlip}:
\begin{eqnarray}
&&N(r)= \gamma r~,
\nonumber\\
&&K(r)= \sqrt{ -\Lambda r^2+\frac{Q^2}{4 r^2}}~.
\label{chargedBTZ2}
\end{eqnarray}
As we observe, the obtained solutions correspond to a
``deformed'', charged BTZ-like solution, and they completely coincide with
it in the limit $f(T)\rightarrow0$ (that is when  $\alpha\rightarrow0$). 
Once again we stress that the above deformation is not of a trivial type,
since we obtain qualitatively different novel terms, which was not
the case in the pure gravitational solutions of the previous section.

The most important qualitative difference is that although in the case of
charged BTZ-like solution (\ref{chargedBTZ2}) there is always a horizon at
$r_H=\left(\frac{Q^2}{4\Lambda}\right)^{1/4}$ that shields the physical
singularity at $r_s=0$, in our case, apart from the positive branch of UV
modification,  we obtained naked physical singularities at $r_s=0$. Furthermore, as expected, in the zero-electric-charge limit we re-obtain the
results of the previous section. 

Finally, in the asymptotic region
($r\to \infty$), the above solutions exhibit the behavior
\begin{equation}
\label{noradial3}
K\left( r\right) \to \sqrt{\frac{-1\pm \sqrt{1-24\alpha \Lambda }}{%
12\alpha }}\, r~,
\end{equation} 
for the UV modification, and 
\begin{equation}  
K\left( r\right) \to \sqrt{-\frac{\Lambda }{2}\pm \frac{\sqrt{%
3\alpha +\Lambda ^{2}}}{2}}\, r~,
\end{equation}
for the IR modification, which coincides with the asymptotic behavior of 
the charged BTZ-like solution (\ref{chargedBTZ2}) with an effective
cosmological constant.   

For completeness we close this section by mentioning an interesting
feature of the 3D $f(T)$-Maxwell theory at hand, namely that it accepts
AdS $pp$-wave solutions
\cite{Peres:1959mm,Skenderis:2002vf,AyonBeato:2004fq}. The relevant
calculations are shown in the Appendix \ref{pp}.   

\section{Final Remarks}
\label{conclusions}

In this work we presented teleparallel gravity in three
dimensions and we examined its circularly symmetric solutions.
Furthermore, we extended our analysis considering functions $f(T)$ of the
torsion scalar, that is formulating 3D $f(T)$ gravity, and we examined the
circularly symmetric solutions too. Finally, we extended our analysis
taking into account the electromagnetic sector, in order to extract the
charged circularly symmetric solutions.

In the simple case of teleparallel 3D gravity, we showed that for a
negative cosmological constant one can obtain the BTZ solution of standard
3D (GR-like) gravity, while for a positive cosmological constant one
acquires the standard Deser-de-Sitter solution. Such a complete
coincidence between teleparallel 3D gravity and standard 3D 
gravity was expected, since the theory is linear in the torsion scalar $T$
and in this case the equivalence of the above gravitational formulations
is complete in all dimensionalities.

In the case of $f(T)$ 3D gravity, after formulating it for a general
torsion scalar, we showed that one can obtain a BTZ-like
solution corresponding to an effective cosmological constant, even in
the case of the standard torsion scalar definition. In
particular, one obtains an effective cosmological constant which depends
on the initial cosmological constant as well as on the parameters
of the used $f(T)$ ansatz. Moreover, we saw that a negative cosmological
constant is not
required for such a BTZ-like solution. This is a difference with
standard 3D gravity, and indicates the novel features that the $f(T)$
structure induces in the gravitational theory. Additionally, and in the
same lines, a positive cosmological constant is not required for the
Deser-de-Sitter-like solution. Finally, note
that the circularly symmetric solutions of 3D $f(T)$ gravity are also
different from the corresponding solutions of $f(R)$ gravity in three
dimensions \cite{delaCruzDombriz:2009et}, which was also expected since it
is well known that $f(T)$ and $f(R)$ modified gravitational theories are
quite different.

In the case of Maxwell-$f(T)$ 3D gravity, interestingly enough we found
that the known no-go theorem of standard (GR-like) 3D gravity
\cite{Cataldo:2002fh,Blagojevic00}, which dynamically excludes 
configurations with two non-vanishing components of the Maxwell field,
is valid too, even going beyond the simple diagonal relation between the
metric and the vierbeins. Thus, examining separately the case of radial or
azimuthal electric field, and considering UV and IR $f(T)$ modifications of
3D gravity, we showed that the theory accepts ``deformed'' charged BTZ-like
solutions, which coincide with the exact standard 3D  result in the
limit $f(T)\rightarrow0$. Moreover, contrary to the simple $f(T)$ case
where the torsion scalar $T$ was a constant, in the Maxwell-$f(T)$ case $T$
has in general an $r$-dependence, a behavior that reveals the new
features brought in by the richer structure of the addition of the
electromagnetic sector. 

However, the most interesting feature of the 3D $f(T)$-Maxwell theory is
that the deformation of the standard charged BTZ solution is not of a
trivial type, since we obtain qualitatively different novel terms and
radically different behavior, contrary to the pure gravitational solutions
where the deformation is expressed only through changes in the
coefficients. In particular, we analyzed the singularities and the horizons
of specific (but quite general) numerical examples. 
Although one can find Killing horizons that shield the physical
singularities, in the majority of the examined cases there are always
parameter choices that lead to the appearance of naked singularities.
This violation of cosmic censorship, that disappears only in the limit
$f(T)\rightarrow0$, may serve as another disadvantage of the $f(T)$
extension of teleparallel gravity, although charged BTZ black holes could
also exhibit naked singularities under special conditions
\cite{Park:1999nc,Hendi:2010px}.
Moreover, we examined the asymptotic behavior of the solutions far away
from the circular object, comparing it with the corresponding behavior of
the usual charged BTZ-like solution (\ref{chargedBTZ2}). In summary, the
novel obtained behavior of the $f(T)$-Maxwell theory reveals the new
features that the $f(T)$ structure brings in 3D gravity. Finally, for
completeness we showed that this theory supports AdS $pp$-wave solutions.
 
In conclusion, the analysis of the present work can be enlightening both
for 3D gravity, since the new features that are brought in by the
$f(T)$ structure may contribute to its quantization efforts, as well as for
$f(T)$ structure itself, since it may bring light to the Lorentz
invariance issues that appear in 4D.\\

\vskip .2in \noindent {\large{{\bf {Acknowledgments}}}}

We wish to thank K. Bamba, C. Boehmer, Y-F. Cai, S. Capozziello,  S.-H.
Chen,  J. B. Dent, S. Dutta R. Ferraro, F. Fiorini, A.
Kehagias, M. Li, J. W. Maluf, 	J. G. Pereira, M. E. Rodrigues, T.
Sotiriou, T. Wang and H. Wei for useful discussions, and an anonymous
referee for useful comments.
Y.V. was supported by Direcci\'{o}n de Investigaci\'{o}n y Desarrollo,
Universidad de la Frontera, DIUFRO DI11-0071.\\

\appendix
\section{No-go theorem in 3D Maxwell-$f(T)$ gravity for non-diagonal
vierbein choice}
\label{nogogeneral}

Let us go beyond the simple diagonal vierbein choice
(\ref{BTZvierbeins2}) and consider a non-diagonal ansatz corresponding to
the same metric (\ref{metricc}). As we have said, this new vierbein choice
will arise from a Lorentz transformation of the diagonal one, namely:
\begin{equation}
e^{a^{\prime }}=\Lambda _{\  a}^{a^{\prime }}e^{a}~,
\end{equation}
denoting the new indices using primes.
Without loss of generality we consider a boost transformation of the form
\footnote{According to \cite{fTLorinv2} D-dimensional $f(T)$ gravity has
$D-1$ new degrees of freedom, which is an indication that they will
correspond to boosts instead of rotations. This can be verified by 
 a detailed investigation of the general vierbein
choice in 3D and 4D $f(T)$ gravity, and its relation to extra degrees of
freedom \cite{inprep}.}
\begin{equation}
\Lambda _{\  a}^{a^{\prime }}\left( x\right) =\left( 
\begin{array}{ccc}
\cosh \theta & \sinh \theta & 0 \\ 
\sinh \theta & \cosh \theta & 0 \\ 
0 & 0 & 1%
\end{array}%
\right) 
\end{equation}
where $\theta \equiv\theta \left( t,r,\phi\right) $, and we mention that 
 $\Lambda _{a^{\prime }}^{\  a}$ is the
inverse of $\Lambda _{\  a}^{a^{\prime }}$.
Therefore, the new vierbein reads
\begin{eqnarray}
&&e^{0^{\prime }}=N\cosh \theta dt+\frac{1}{K}\sinh \theta dr~,\nonumber\\
&&e^{1^{\prime }}=N\sinh \theta dt+\frac{1}{K}\cosh \theta  dr~, 
\nonumber\\
&&e^{2^{\prime }}=e^{2}~.
\end{eqnarray}
Inserting these into the torsion scalar (\ref{scalartorsion}), with
$\rho_{1}=0$, $\rho_{2}=-\frac{1}{2}$, $\rho_{4}=1$, we find that the
new torsion scalar will be
\begin{equation}
T^{\prime }=T-2\frac{K}{Nr}\frac{\partial \theta }{\partial t}~.
\end{equation}
Similarly,  the electric sector of the electromagnetic 
2-form  (\ref{F}) will be
\begin{equation}
F^{\prime }=E_{r}^{\prime }e^{0^{\prime }}e^{1^{\prime }}+E_{\phi }^{\prime
}e^{2^{\prime }}e^{0^{\prime }}~,
\end{equation}
and correspondingly one can find the energy momentum tensor for the
electromagnetic field $S^a_b$ using (\ref{elenermom}). Therefore, it is easy to see that the variation of the electric part of
the Lagrangian (\ref{Mlagrangian}) will now give
\begin{equation}
\frac{\delta {\cal{L}} _{F}}{\delta e^{a^{\prime }}}=-\epsilon _{a^{\prime
}b^{\prime }c^{\prime }}s^{b^{\prime }}e^{c^{\prime }}~,
\end{equation}
that is one replaces the old quantities by the prime-ones.

So in summary, we can see that the new field equations will have the form
of (\ref{fieldeq}) but with all quantities replaced by the prime-ones.
Obviously, we observe that the equations not involving the electric
fields will be more complicated that those of
(\ref{firsteq})-(\ref{maxwell3}), and thus the solution structure will be
different. However, the contribution of the electric fields is given by 
\begin{eqnarray}
-\epsilon _{1^{\prime }b^{\prime }c^{\prime
}}s^{b^{\prime }}e^{c^{\prime }}&=&\frac{1}{2}\left( E_{r}^{\prime
2}-E_{\phi
}^{\prime 2}\right) e^{0^{\prime }}e^{2^{\prime }}+E_{r}^{\prime }E_{\phi
}^{\prime }e^{0^{\prime }}e^{1^{\prime }}\nonumber\\
&=&\frac{1}{2}\left(
E_{r}^{\prime
2}-E_{\phi }^{\prime 2}\right) \left( \cosh \theta\, e^{0}+\sinh \theta\,
e^{1}\right) e^{2}-E_{r}^{\prime }E_{\phi }^{\prime
}e^{0}e^{1}~,
\end{eqnarray}
where we take $a^{\prime}=1^{\prime}$. Then, by substituting in the
fields equations we find that
\begin{eqnarray}
E_{r}^{\prime }E_{\phi }^{\prime }=0~,
\end{eqnarray}
 that is the no-go theorem (\ref{electric})
is still valid in the general vierbein choice.

\section{$pp$-wave solutions in 3D $f(T)$-Maxwell theory}
\label{pp}

In this appendix we show that the 3D $f(T)$-Maxwell theory accepts the
interesting class of solutions known as AdS $pp$-waves
\cite{Peres:1959mm,Skenderis:2002vf,AyonBeato:2004fq}. The corresponding
metric reads:
\begin{equation}
ds^2=h(y)^2\left[-2H(u,y)du^2-2dudv+dy^2\right]~.
\end{equation}
We consider the triad as
\begin{equation}
\label{ppvireb}
e^0=h(y)\left(\frac{H+1}{2}du+dv\right)~,\quad e^1=h(y)dy~,\quad e^2=h(y)\left(\frac{H-1}{2}du+dv\right)~,
\end{equation}
and the electromagnetic potential as
\begin{equation}
A=a(u,y)du~.
\end{equation}
Then
\begin{equation}
F=dA=-\frac{1}{h^2}\frac{\partial a}{\partial y}e^0\wedge e^1-\frac{1}{h^2}\frac{\partial a}{\partial y}e^1\wedge e^2~,
\end{equation}
and the field equations are given by
\begin{eqnarray}
\label{ppeq1}
&&\left[1+\frac{df}{dT}\right]\left[\frac{1}{h}\frac{\partial}{\partial
y}\left(\frac{1}{h}\frac{\partial H}{\partial
y}\right)-2\frac{h'}{h^3}\frac{\partial H}{\partial
y}\right]+\frac{1}{h^2}\frac{d^2f}{dT^2}\frac{\partial T}{\partial
y}\frac{\partial H}{\partial y}-\left(\frac{1}{h^2}\frac{\partial
a}{\partial y}\right)^2=0~,
 \\
\label{ppeq2}
&&\left[1+\frac{df}{dT}\right]\frac{1}{h}\frac{\partial}{\partial
y}\left(\frac{h'}{h^2}\right)+\frac{d^2f}{dT^2}\frac{\partial T}{\partial
y}\frac{h'}{h^3}=0~,
 \\
\label{ppeq3}
&&\left[1+2\frac{df}{dT}\right]T-f(T)+2\Lambda=0~,
\end{eqnarray}
with $h'=dh(y)/dy$. Using the vierbein choice (\ref{ppvireb}) and
the definition of the torsion scalar (\ref{actiontel3D}) we can calculate
\begin{equation}
T=2\left(\frac{h'}{h^2}\right)^2~.
\end{equation}
Now, using the Maxwell equations we get 
$\frac{\partial}{\partial y}\left(\frac{1}{h}\frac{\partial a}{\partial
y}\right)=0$
and in summary we result to the pp-wave solutions
\begin{eqnarray}
&&h(y)=\frac{1}{y}\sqrt{\frac{2}{T}}~,
\nonumber\\
&&a(u,y)=\sqrt{\frac{2}{T}}k(u)\ln y+j(u)~,
\nonumber\\
&&H(u,y)=\frac{k^2(u)}{8\left[1+\frac{df}{dT}\right]}y^2+g(u)~,
\end{eqnarray}
where $k(u)$ and $j(u)$ are arbitrary function and the scalar torsion is constant.
Finally, note that in  the special case where 
$f(T)= -T+2\Lambda+\sqrt{T}$, equation (\ref{ppeq3}) is satisfied
identically and thus the torsion scalar is not restricted to be a
constant. Equation (\ref{ppeq2})  is satisfied too, and therefore
from (\ref{ppeq1}) we obtain
\begin{eqnarray}
&&H(u,y)= h^2(y)k(u)+g(u)~,
\nonumber\\
&&a(u,y)=j(u)~,
\end{eqnarray}
with $h(y)$, $k(u)$, $g(u)$ and $j(u)$  arbitrary
functions.

\providecommand{\href}[2]{#2}

\begingroup

\raggedright

\endgroup


\begin{thebibliography}{10}



\bibitem{BTZ}
 M.~Banados, C.~Teitelboim and J.~Zanelli,
 {\it {The Black hole in
three-dimensional space-time}},
 Phys.\ Rev.\ Lett.\ \textbf{69}, 1849 (1992),
[\href{http://xxx.lanl.gov/abs/hep-th/9204099]}
{{\tt arXiv:hep-th/9204099]}}].


\bibitem{Carlip} 
S.~Carlip, 
{\it{The (2+1)-Dimensional black hole}},
 Class.\
Quant.\ Grav.\ \textbf{12}, 2853 (1995),
[\href{http://xxx.lanl.gov/abs/gr-qc/9506079}
{{\tt arXiv:gr-qc/9506079}}].

\bibitem{Carlip:2005zn}
  S.~Carlip,
{\it {Conformal field theory, (2+1)-dimensional gravity, and the BTZ black
hole}},
  Class.\ Quant.\ Grav.\  {\bf 22}, R85-R124 (2005),
  [\href{http://xxx.lanl.gov/abs/gr-qc/0503022}
{{\tt arXiv:gr-qc/0503022}}].



\bibitem{deser}
  S.~Deser, R.~Jackiw, S.~Templeton,
{\it{Topologically Massive Gauge Theories}}
  Annals Phys.\  {\bf 140}, 372 (1982).

\bibitem{deser1}
  S.~Deser, R.~Jackiw, S.~Templeton,
{\it{Three-Dimensional Massive Gauge Theories}},
  Phys.\ Rev.\ Lett.\  {\bf 48}, 975 (1982).

\bibitem{Li:2008dq} 
 W.~Li, W.~Song and A.~Strominger,  
{\it{Chiral Gravity in Three Dimensions}},
  JHEP \textbf{0804}, 082 (2008) 
[\href{http://xxx.lanl.gov/abs/0801.4566}
{{\tt arXiv:0801.4566}}].
 
 

\bibitem{Strominger:2008dp} 
 A.~Strominger,
{\it{A Simple Proof of the Chiral
Gravity Conjecture}},
[\href{http://xxx.lanl.gov/abs/0808.0506}
{{\tt arXiv:0808.0506}}].


\bibitem{Carlip:2008jk}  
S.~Carlip, S.~Deser, A.~Waldron and D.~K.~Wise, 
{\it{Cosmological Topologically Massive Gravitons and Photons}}, 
Class.\
Quant.\ Grav.\ \textbf{26}, 075008 (2009),
[\href{http://xxx.lanl.gov/abs/0803.3998}
{{\tt arXiv:0803.3998}}].
 

\bibitem{Carlip:2008eq} 
 S.~Carlip, S.~Deser, A.~Waldron and D.~K.~Wise, 
{\it{Topologically Massive AdS Gravity}},
 Phys.\ Lett.\ B \textbf{666}, 272
(2008),
 [\href{http://xxx.lanl.gov/abs/0807.0486}
{{\tt arXiv:0807.0486}}].
 

\bibitem{Carlip:2008qh}  
S.~Carlip, 
 {\it{The Constraint Algebra of
Topologically Massive AdS Gravity}}, 
JHEP \textbf{0810}, 078 (2008), [\href{http://xxx.lanl.gov/abs/0807.4152}
{{\tt arXiv:0807.4152}}].
 

\bibitem{Giribet:2008bw} 
 G.~Giribet, M.~Kleban and M.~Porrati, 
{\it{Topologically Massive Gravity at the Chiral Point is Not Chiral}},
JHEP 
\textbf{0810}, 045 (2008),
[\href{http://xxx.lanl.gov/abs/0807.4703}
{{\tt arXiv:0807.4703}}].
 

\bibitem{Park:2008yy} 
 M.~i.~Park,
{\it{Constraint Dynamics and Gravitons in
Three Dimensions}},
  JHEP \textbf{0809}, 084 (2008),
[\href{http://xxx.lanl.gov/abs/0805.4328}
{{\tt arXiv:0805.4328}}].
 

\bibitem{Blagojevic:2008bn}
  M.~Blagojevic and B.~Cvetkovic, 
{\it{Canonical
structure of topologically massive gravity with a cosmological 
constant}}, 
JHEP \textbf{0905}, 073 (2009),
[\href{http://xxx.lanl.gov/abs/0812.4742}
{{\tt arXiv:0812.4742}}].
 


\bibitem{Grumiller:2008pr} 
 D.~Grumiller, R.~Jackiw and N.~Johansson, 
{\it{Canonical analysis of cosmological topologically massive gravity at
the chiral point}},
[\href{http://xxx.lanl.gov/abs/0806.4185}
{{\tt arXiv:0806.4185}}].
 


\bibitem{Garbarz:2008qn} 
 A.~Garbarz, G.~Giribet and Y.~Vasquez, 
{\it{Asymptotically AdS$_3$ Solutions to Topologically Massive Gravity at
Special  Values of the Coupling Constants}},
  Phys.\ Rev.\  {\bf D79}, 044036 (2009),
[\href{http://xxx.lanl.gov/abs/0811.4464}
{{\tt arXiv:0811.4464}}].
 

\bibitem{Grumiller:2008qz} 
 D.~Grumiller and N.~Johansson, 
{\it{Instability in cosmological topologically massive gravity at the
chiral point}},
 JHEP 
\textbf{0807}, 134 (2008),
[\href{http://xxx.lanl.gov/abs/0805.2610}
{{\tt arXiv:0805.2610}}].
 

\bibitem{Grumiller:2008es}
  D.~Grumiller and N.~Johansson, 
{\it{Consistent boundary conditions for cosmological topologically massive 
gravity at the chiral point}}, 
 Int.\ J.\ Mod.\ Phys.\ D \textbf{17}, 2367 (2009),
[\href{http://xxx.lanl.gov/abs/0808.2575}
{{\tt arXiv:0808.2575}}].
 

\bibitem{Henneaux:2009pw} 
 M.~Henneaux, C.~Martinez and R.~Troncoso, 
{\it{Asymptotically anti-de Sitter spacetimes in topologically massive 
gravity}},
  Phys.\ Rev.\ D \textbf{79}, 081502 (2009), 
[\href{http://xxx.lanl.gov/abs/0901.2874}
{{\tt arXiv:0901.2874}}].

\bibitem{Maloney:2009ck}
  A.~Maloney, W.~Song and A.~Strominger,
{\it{Chiral Gravity, Log Gravity and Extremal CFT}},
 Phys.\ Rev.\ D \textbf{81}, 064007 (2010),
[\href{http://xxx.lanl.gov/abs/0903.4573}
{{\tt arXiv:0903.4573}}].
 

\bibitem{AyonBeato:2009nh} 
 E.~Ayon-Beato, A.~Garbarz, G.~Giribet and
M.~Hassaine, 
{\it{Lifshitz Black Hole in Three Dimensions}},
 Phys.\ Rev.\ D 
\textbf{80}, 104029 (2009),
[\href{http://xxx.lanl.gov/abs/0909.1347}
{{\tt arXiv:0909.1347}}]. 
 

\bibitem{ein28}  
 A.~Unzicker and T.~Case,  
{\it{Translation of Einstein's attempt of a unified field
theory with teleparallelism}},
[\href{http://xxx.lanl.gov/abs/physics/0503046}
{{\tt arXiv:physics/0503046}}].



\bibitem{Hayashi79}  
K. Hayashi and T. Shirafuji, 
{\it{New General Relativity}},
 Phys. Rev. D \textbf{19}, 3524 (1979), [{\it{Addendum-ibid}}.
D \textbf{24}, 3312 (1982)].





\bibitem{Kawai1} 
  T.~Kawai,
{\it {Teleparallel theory of (2+ 1)-dimensional gravity}},
  Phys.\ Rev.\ D {\bf 48}, 5668 (1993).


\bibitem{Kawai2} 
     T.~Kawai,
{\it {Exotic black hole solution in teleparallel theory of (2+1)
dimensional gravity}},
  Prog.\ Theor.\ Phys.\  {\bf 94}, 1169 (1995),
  [\href{http://xxx.lanl.gov/abs/gr-qc/9410032}
{{\tt arXiv:gr-qc/9410032}}].



\bibitem{Kawai3} 
  T.~Kawai,
{\it {Generators of internal Lorentz transformations and of general affine
 coordinate transformations in teleparallel theory of (2+1)-dimensional
 gravity. Cases with static circularly symmetric space-times}},
  Prog.\ Theor.\ Phys.\  {\bf 94}, 915 (1995),
  [\href{http://xxx.lanl.gov/abs/gr-qc/9507017}
{{\tt arXiv:gr-qc/9507017}}].



\bibitem{3dgravitywithtorsion}
 E.~W.~Mielke and P.~Baekler,
{\it{Topological Gauge Model Of
Gravity With Torsion}},
 Phys.\ Lett.\ A \textbf{156}, 399 (1991).

\bibitem{Sousa:2003sx}
  A.~A.~Sousa, J.~W.~Maluf,
  {\it{Black holes in 2+1 teleparallel theories of gravity}},
  Prog.\ Theor.\ Phys.\  {\bf 108}, 457-470 (2002),
[\href{http://xxx.lanl.gov/abs/gr-qc/0301079}
{{\tt arXiv:gr-qc/0301079}}].


\bibitem{garcia} 
A.~A.~Garcia, F.~W.~Hehl, C.~Heinicke and A.~Macias,
{\it{Exact vacuum solution of a (1+2)-dimensional Poincare gauge theory:
BTZ solution with torsion}},
Phys.\ Rev.\ D \textbf{67}, 124016 (2003),
[\href{http://xxx.lanl.gov/abs/gr-qc/0302097}
{{\tt arXiv:gr-qc/0302097}}].
 


\bibitem{Blagojevic00}
 M.~Blagojevic and B.~Cvetkovic, 
{\it{Electric field in 3D gravity with torsion}},
 Phys.\ Rev.\ D \textbf{78}, 044036 (2008),
[\href{http://xxx.lanl.gov/abs/0804.1899}
{{\tt arXiv:0804.1899}}].
 

\bibitem{Blagojevic11}
 M.~Blagojevic and B.~Cvetkovic, 
{\it{Self-dual Maxwell field in 3D gravity with torsion}},
 Phys.\ Rev.\ D \textbf{78}, 044037
(2008),
[\href{http://xxx.lanl.gov/abs/0805.3627}
{{\tt arXiv:0805.3627}}].
 


\bibitem{Blagojevic22}
  M.~Blagojevic, B.~Cvetkovic and O.~Miskovic,
{\it{Nonlinear electrodynamics in 3D gravity with torsion}},
  Phys.\ Rev.\  D {\bf 80}, 024043 (2009),
[\href{http://xxx.lanl.gov/abs/0906.0235}
{{\tt arXiv:0906.0235}}].


\bibitem{Vasquez:2009mk}
 P.~ A.~Gonzalez and Y.~Vasquez,
{\it{Exact solutions in 3D gravity with torsion}},
JHEP {\bf 1108}, 089 (2011),
[\href{http://xxx.lanl.gov/abs/0907.4165}
{{\tt arXiv:0907.4165}}].
 

\bibitem{Santamaria:2011cz} 
  R.~C.~Santamaria, J.~D.~Edelstein, A.~Garbarz and G.~E.~Giribet,
{\it{On the addition of torsion to chiral gravity}},
  Phys.\ Rev.\ D {\bf 83}, 124032 (2011),
[\href{http://xxx.lanl.gov/abs/1102.4649}
{{\tt arXiv:1102.4649}}].
 
\bibitem{fT}
R.~Ferraro and F.~Fiorini,
  {\it{Modified teleparallel gravity: Inflation without inflaton}},
Phys.\ Rev.\  D {\bf 75} 084031 (2007), 
 [\href{http://xxx.lanl.gov/abs/gr-qc/0610067}
{{\tt arXiv:gr-qc/0610067}}].

\bibitem{Ferraro:2008ey}
  R.~Ferraro, F.~Fiorini,
  {\it{On Born-Infeld Gravity in Weitzenbock spacetime}},
  Phys.\ Rev.\  {\bf D78}, 124019 (2008),
   [\href{http://xxx.lanl.gov/abs/0812.1981}
{{\tt arXiv:0812.1981}}].
    
\bibitem{Bengochea:2008gz}
  G.~R.~Bengochea and R.~Ferraro,
  {\it{Dark torsion as the cosmic speed-up}},
  Phys.\ Rev.\ D \textbf{79}, 124019 (2009),
[\href{http://xxx.lanl.gov/abs/0812.1205}
{{\tt arXiv:0812.1205}}].
 
\bibitem{Linder:2010py}
  E.~V.~Linder,
  {\it{Einstein's Other Gravity and the Acceleration of the
Universe}},
  Phys.\ Rev.\ D \textbf{81}, 127301 (2010),
[\href{http://xxx.lanl.gov/abs/1005.3039}
{{\tt arXiv:1005.3039}}].
 
\bibitem{Myrzakulov:2010vz}
  R.~Myrzakulov,
  {\it{Accelerating universe from F(T) gravities}},
[\href{http://xxx.lanl.gov/abs/1006.1120}
{{\tt arXiv:1006.1120}}].

\bibitem{Chen001}
  S.~H.~Chen, J.~B.~Dent, S.~Dutta and E.~N.~Saridakis,
  {\it{Cosmological perturbations in f(T) gravity}},
  Phys.\ Rev.\ D \textbf{83}, 023508 (2011),
[\href{http://xxx.lanl.gov/abs/1008.1250}
{{\tt arXiv:1008.1250}}].

\bibitem{Wu001}
  P.~Wu, H.~W.~Yu,
  {\it{$f(T)$ models with phantom divide line crossing}},
  Eur.\ Phys.\ J.\ \textbf{C71}, 1552 (2011),
[\href{http://xxx.lanl.gov/abs/1008.3669}
{{\tt arXiv:1008.3669}}].
 

\bibitem{Bamba:2010iw}
  K.~Bamba, C.~-Q.~Geng, C.~-C.~Lee,
  {\it{Comment on 'Einstein's Other Gravity and the Acceleration of
the Universe}},
[\href{http://xxx.lanl.gov/abs/1008.4036}
{{\tt arXiv:1008.4036}}].


\bibitem{Dent001}
J.~B.~Dent, S.~Dutta, E.~N.~Saridakis,
  {\it{f(T) gravity mimicking dynamical dark energy. Background and
perturbation analysis}},
  JCAP {\bf 1101}, 009 (2011),
[\href{http://xxx.lanl.gov/abs/1010.2215}
{{\tt arXiv:1010.2215}}].



\bibitem{Zheng:2010am}
  R.~Zheng, Q.~-G.~Huang,
  {\it{Growth factor in $f(T)$ gravity}},
  JCAP \textbf{1103}, 002 (2011),
[\href{http://xxx.lanl.gov/abs/1010.3512}
{{\tt arXiv:1010.3512}}].


\bibitem{Bamba:2010wb}
  K.~Bamba, C.~-Q.~Geng, C.~-C.~Lee, L.~-W.~Luo,
  {\it{Equation of state for dark energy in $f(T)$ gravity}},
  JCAP \textbf{1101}, 021 (2011),
[\href{http://xxx.lanl.gov/abs/1011.0508}
{{\tt arXiv:1011.0508}}].



 

\bibitem{Yerzhanov:2010vu}
  K.~K.~Yerzhanov, S.~R.~Myrzakul, I.~I.~Kulnazarov and
R.~Myrzakulov,
  {\it{Accelerating cosmology in F(T) gravity with scalar field}},
[\href{http://xxx.lanl.gov/abs/1006.3879}
{{\tt arXiv:1006.3879}}].


\bibitem{Yang:2010ji}
  R.~-J.~Yang,
  {\it{Conformal transformation in $f(T)$ theories}},
  Europhys.\ Lett.\ \textbf{93}, 60001 (2011),
[\href{http://xxx.lanl.gov/abs/1010.1376}
{{\tt arXiv:1010.1376}}].


\bibitem{Wu:2010mn}
  P.~Wu, H.~W.~Yu,
  {\it{Observational constraints on $f(T)$ theory}},
  Phys.\ Lett.\ \textbf{B693}, 415 (2010),
[\href{http://xxx.lanl.gov/abs/1006.0674}
{{\tt arXiv:1006.0674}}].

\bibitem{Bengochea001}
    G.~R.~Bengochea,
  {\it{Observational information for f(T) theories and Dark Torsion}},
  Phys.\ Lett.\  {\bf B695}, 405 (2011),
[\href{http://xxx.lanl.gov/abs/1008.3188}
{{\tt arXiv:1008.3188}}].


\bibitem{Wu:2010xk}
  P.~Wu, H.~W.~Yu,
  {\it{The dynamical behavior of $f(T)$ theory}},
  Phys.\ Lett.\ \textbf{B692}, 176 (2010),
[\href{http://xxx.lanl.gov/abs/1007.2348}
{{\tt arXiv:1007.2348}}].


\bibitem{Zhang:2011qp}
   Y.~Zhang, H.~Li, Y.~Gong, Z.~-H.~Zhu,
  {\it{Notes on $f(T)$ Theories}},
  JCAP {\bf 1107}, 015 (2011),
[\href{http://xxx.lanl.gov/abs/1103.0719}
{{\tt arXiv:1103.0719}}].

\bibitem{Ferraro001}
   R.~Ferraro, F.~Fiorini,
  {\it{Non trivial frames for f(T) theories of gravity and beyond}},
  Phys.\ Lett.\  {\bf B702}, 75 (2011),
[\href{http://xxx.lanl.gov/abs/1103.0824}
{{\tt arXiv:1103.0824}}].


\bibitem{Cai:2011tc}
  Y.~-F.~Cai, S.~-H.~Chen, J.~B.~Dent, S.~Dutta, E.~N.~Saridakis,
  {\it{Matter Bounce Cosmology with the f(T) Gravity}},
  Class.\ Quant.\ Grav.\  {\bf 28}, 2150011 (2011),
[\href{http://xxx.lanl.gov/abs/1104.4349}
{{\tt arXiv:1104.4349}}].
  [arXiv:1104.4349 [astro-ph.CO]].


\bibitem{Chattopadhyay001}
  S.~Chattopadhyay, U.~Debnath,
  {\it{Emergent universe in chameleon, f(R) and f(T) gravity theories}},
  Int.\ J.\ Mod.\ Phys.\  {\bf D20}, 1135 (2011),
[\href{http://xxx.lanl.gov/abs/1105.1091}
{{\tt arXiv:1105.1091}}].

\bibitem{Sharif001}
  M.~Sharif, S.~Rani,
  {\it{F(T) Models within Bianchi Type I Universe}},
  Mod.\ Phys.\ Lett.\  {\bf A26}, 1657 (2011),
[\href{http://xxx.lanl.gov/abs/1105.6228}
{{\tt arXiv:1105.6228}}].

\bibitem{Wei001}
  H.~Wei, X.~-P.~Ma, H.~-Y.~Qi,
{\it{$f(T)$ Theories and Varying Fine Structure Constant}}
  Phys.\ Lett.\  {\bf B703}, 74 (2011),
 [\href{http://xxx.lanl.gov/abs/1106.0102}
{{\tt arXiv:1106.0102}}].


\bibitem{Ferraro002}
  R.~Ferraro, F.~Fiorini,
  {\it{Cosmological frames for theories with absolute parallelism}},
[\href{http://xxx.lanl.gov/abs/1106.6349}
{{\tt arXiv:1106.6349}}].
 


\bibitem{Boehmer004}
  C.~G.~Boehmer, A.~Mussa, N.~Tamanini,
  {\it{Existence of relativistic stars in f(T) gravity}},
[\href{http://xxx.lanl.gov/abs/1107.4455}
{{\tt arXiv:1107.4455}}].
 



\bibitem{Wei005}
  H.~Wei, H.~-Y.~Qi, X.~-P.~Ma,
  {\it{Constraining $f(T)$ Theories with the Varying Gravitational
Constant}},
 [\href{http://xxx.lanl.gov/abs/1108.0859}
{{\tt arXiv:1108.0859}}].
  

\bibitem{Capozziello006}
  S.~Capozziello, V.~F.~Cardone, H.~Farajollahi, A.~Ravanpak,
  {\it{Cosmography in f(T)-gravity}},
[\href{http://xxx.lanl.gov/abs/1108.2789}
{{\tt arXiv:1108.2789}}].
 

\bibitem{Wu007}
  P.~Wu, H.~Yu,
  {\it{The stability of the Einstein static state in $f(T)$ gravity}},
  Phys.\ Lett.\  {\bf B703}, 223 (2011), 
[\href{http://xxx.lanl.gov/abs/1108.5908}
{{\tt arXiv:1108.5908}}].
 

\bibitem{Daouda001}
  M.~H.~Daouda, M.~E.~Rodrigues, M.~J.~S.~Houndjo,
  {\it{Static Anisotropic Solutions in $f(T)$ Theory}},
[\href{http://xxx.lanl.gov/abs/1109.0528}
{{\tt arXiv:1109.0528}}].


\bibitem{Daouda:2011iy}
  M.~H.~Daouda, M.~E.~Rodrigues, M.~J.~S.~Houndjo,
{\it {New Static Solutions in f(T) Theory}},
  [\href{http://xxx.lanl.gov/abs/1108.2920}
{{\tt arXiv:1108.2920}}].
 


\bibitem{Bamba:2011pz}
  K.~Bamba, C.~-Q.~Geng,
  {\it{Thermodynamics of cosmological horizons in $f(T)$ gravity}},
[\href{http://xxx.lanl.gov/abs/1109.1694}
{{\tt arXiv:1109.1694}}].
 
\bibitem{Geng:2011aj}
  C.~-Q.~Geng, C.~-C.~Lee, E.~N.~Saridakis, Y.~-P.~Wu,
  {\it{`Teleparallel' Dark Energy}},
  Phys.\ Lett.\  {\bf B704}, 384-387 (2011)'
[\href{http://xxx.lanl.gov/abs/1109.1092}
{{\tt arXiv:1109.1092}}].

\bibitem{Wei:2011yr}
  H.~Wei,
  {\it{Dynamics of Teleparallel Dark Energy}},
  [\href{http://xxx.lanl.gov/abs/1109.6107}
{{\tt arXiv:1109.6107}}].


\bibitem{Geng:2011ka}
  C.~-Q.~Geng, C.~-C.~Lee, E.~N.~Saridakis,
 {\it{Observational Constraints on Teleparallel Dark Energy}},
  JCAP {\bf 1201}, 002 (2012),
[\href{http://xxx.lanl.gov/abs/1110.0913}
{{\tt arXiv:1110.0913}}].
  
\bibitem{Xu:2012jf} 
  C.~Xu, E.~N.~Saridakis and G.~Leon,
{\it {Phase-Space analysis of Teleparallel Dark Energy}},
  [\href{http://xxx.lanl.gov/abs/1202.3781}
{{\tt arXiv:1202.3781}}].

\bibitem{Iorio:2012cm} 
  L.~Iorio and E.~N.~Saridakis,
{\it {Solar system constraints on f(T) gravity}},
  [\href{http://xxx.lanl.gov/abs/1203.5781}
{{\tt arXiv:1203.5781}}].
 

\bibitem{Wang:2011xf}
    T.~Wang,
  {\it{Static Solutions with Spherical Symmetry in f(T) Theories}},
  Phys.\ Rev.\  {\bf D84}, 024042 (2011),
[\href{http://xxx.lanl.gov/abs/1102.4410}
{{\tt arXiv:1102.4410}}].

\bibitem{Miao003}
  R.~-X.~Miao, M.~Li, Y.~-G.~Miao,
  {\it{Violation of the first law of black hole thermodynamics in $f(T)$
gravity}}, 
[\href{http://xxx.lanl.gov/abs/1107.0515}
{{\tt arXiv:1107.0515}}].
  
 
\bibitem{Wei:2011aa} 
  H.~Wei, X.~-J.~Guo and L.~-F.~Wang,
{\it {Noether Symmetry in $f(T)$ Theory}},
  Phys.\ Lett.\ B {\bf 707}, 298 (2012),
  [\href{http://xxx.lanl.gov/abs/1112.2270}
{{\tt arXiv:1112.2270}}].

 



\bibitem{Ferraro:2011ks}
  R.~Ferraro, F.~Fiorini,
{\it {Spherically symmetric static spacetimes in vacuum f(T) gravity}},
  Phys.\ Rev.\ D {\bf 84}, 083518 (2011),
  [\href{http://xxx.lanl.gov/abs/1109.4209}
{{\tt arXiv:1109.4209}}].
 


 

 

\bibitem{Weitzenb23}  Weitzenb\"{o}ck R.,  {\it{Invarianten Theorie}}, 
Nordhoff, Groningen, The Netherlands (1923).



\bibitem{Maluf:1994ji}
  J.~W.~Maluf,  
{\it{Hamiltonian formulation of the teleparallel description of
general relativity}},
J.\ Math.\ Phys.\ \textbf{35} (1994) 335.



\bibitem{Arcos:2005ec} 
 H.~I.~Arcos and J.~G.~Pereira,  
{\it{Torsion Gravity: a Reappraisal}},
Int.\ J.\ Mod.\ Phys.\ D \textbf{13}, 2193 (2004),
[\href{http://xxx.lanl.gov/abs/gr-qc/0501017}
{{\tt arXiv:gr-qc/0501017}}].
 
 
\bibitem{Weinberg:2008} 
 S. Weinberg, {\it{Cosmology}},  Oxford University
Press, Oxford U.K., (2008).


\bibitem{Muench:1998ay} 
 U.~Muench, F.~Gronwald and F.~W.~Hehl,  
{\it{A Small guide to variations in teleparallel gauge theories of gravity
and the Kaniel-Itin model}},
Gen.\ Rel.\ Grav.\ \textbf{30}, 933 (1998),
[\href{http://xxx.lanl.gov/abs/gr-qc/9801036}
{{\tt arXiv:gr-qc/9801036}}].
 

\bibitem{Itin:1999wi}
  Y.~Itin,
  {\it{Coframe teleparallel models of gravity: Exact solutions}},
  Int.\ J.\ Mod.\ Phys.\  {\bf D10}, 547-573 (2001),
[\href{http://xxx.lanl.gov/abs/gr-qc/9912013}
{{\tt arXiv:gr-qc/9912013}}].
 

\bibitem{Deser:1983tn}
  S.~Deser, R.~Jackiw, G.~'t Hooft,
{\it{Three-Dimensional Einstein Gravity: Dynamics of Flat Space}},
  Annals Phys.\  {\bf 152}, 220 (1984).
  
\bibitem{fTLorinv0}
   T.~P.~Sotiriou, B.~Li, J.~D.~Barrow,
  {\it{Generalizations of teleparallel gravity and local Lorentz
symmetry}},
  Phys.\ Rev.\  {\bf D83}, 104030 (2011),
[\href{http://xxx.lanl.gov/abs/1012.4039}
{{\tt arXiv:1012.4039}}].

 



\bibitem{fTLorinv1}
  B.~Li, T.~P.~Sotiriou, J.~D.~Barrow,
{\it {Large-scale Structure in f(T) Gravity}},
  Phys.\ Rev.\  {\bf D83}, 104017 (2011),
[\href{http://xxx.lanl.gov/abs/:1103.2786}
{{\tt arXiv:1103.2786}}].


\bibitem{fTLorinv2}
  M.~Li, R.~X.~Miao, Y.~G.~Miao,
{\it {Degrees of freedom of $f(T)$ gravity}},
  JHEP {\bf 1107}, 108 (2011),
[\href{http://xxx.lanl.gov/abs/1105.5934}
{{\tt arXiv:1105.5934}}].


\bibitem{inprep}
  Y.~-F.~Cai, S.~-H.~Chen, J.~B.~Dent, S.~Dutta, E.~N.~Saridakis, in
preparation.



\bibitem{Cataldo:2002fh}
  M.~Cataldo,
{\it {Azimuthal electric field in a static rotationally symmetric
(2+1)-dimensional space-time}},
  Phys.\ Lett.\  {\bf B529}, 143-149 (2002),
[\href{http://xxx.lanl.gov/abs/gr-qc/0201047}
{{\tt arXiv:gr-qc/0201047}}].


\bibitem{Painleve}
P. Painlev\'{e},{\it { La M\'{e}canique classique et la th\'{e}orie de la
relativit\'{e}}}, C. R. Acad. Sci. (Paris), {\bf{173}}, 677-680 
 (1921).

\bibitem{Gullstrand}
A. Gullstrand, {\it {Allegemeine L$\ddot{o}$sung des statischen
Eink$\ddot{o}$rper-problems in der Einsteinschen Gravitations theorie}},
Arkiv. Mat. Astron. Fys. {\bf{16}}(8), 1-15 (1922).

\bibitem{Martel:2000rn} 
  K.~Martel and E.~Poisson, 
 {\it {Regular coordinate systems for Schwarzschild and other
spherical space-times}}, 
  Am.\ J.\ Phys.\  {\bf 69}, 476 (2001),
[\href{http://xxx.lanl.gov/abs/gr-qc/0001069}
{{\tt arXiv:gr-qc/0001069}}].

 

\bibitem{Liu:2005hj} 
  W.~Liu, 
 {\it {New coordinates for BTZ black hole and Hawking radiation via
tunnelling}},
  Phys.\ Lett.\ B {\bf 634}, 541 (2006),
[\href{http://xxx.lanl.gov/abs/gr-qc/0512099}
{{\tt arXiv:gr-qc/0512099}}].

 
  
\bibitem{Hilbert} R. Courant and D. Hilbert,  {\it{Methods of Mathematical
Physics}}, Vol. 2, Cambridge University Press, Cambridge U.K., (1966).



 
  
\bibitem{Peres:1959mm}
  A.~Peres,
{\it {Some Gravitational Waves}},
  Phys.\ Rev.\ Lett.\  {\bf 3}, 571 (1959).
 
\bibitem{Skenderis:2002vf}
  K.~Skenderis, M.~Taylor,
{\it {Branes in AdS and p p wave space-times}},
  JHEP {\bf 0206}, 025 (2002),
  [\href{http://xxx.lanl.gov/abs/hep-th/0204054}
{{\tt arXiv:hep-th/0204054}}].



\bibitem{AyonBeato:2004fq}
  E.~Ayon-Beato, M.~Hassaine,
{\it {pp waves of conformal gravity with self-interacting source}},
  Annals Phys.\  {\bf 317}, 175-181 (2005),
  [\href{http://xxx.lanl.gov/abs/hep-th/0409150}
{{\tt arXiv:hep-th/0409150}}].

\bibitem{delaCruzDombriz:2009et}
  A.~de la Cruz-Dombriz, A.~Dobado, A.~L.~Maroto,
{\it {Black Holes in f(R) theories}},
  Phys.\ Rev.\  {\bf D80}, 124011 (2009),
  [\href{http://xxx.lanl.gov/abs/0907.3872}
{{\tt arXiv:0907.3872}}].





 
 
\bibitem{Park:1999nc} 
  D.~H.~Park and S.~H.~Yang,
 {\it {Geodesic motions in (2+1)-dimensional charged black holes}},
  Gen.\ Rel.\ Grav.\  {\bf 31}, 1343 (1999),
[\href{http://xxx.lanl.gov/abs/gr-qc/9901027}
{{\tt arXiv:gr-qc/9901027}}].


 

\bibitem{Hendi:2010px} 
  S.~H.~Hendi,
  {\it {Charged BTZ-like Black Holes in Higher Dimensions}},
  Eur.\ Phys.\ J.\ C {\bf 71}, 1551 (2011),
[\href{http://xxx.lanl.gov/abs/1007.2704}
{{\tt arXiv:1007.2704}}].


 


 
\end{thebibliography}
\end{document}